\documentclass[]{article}

\usepackage{epstopdf}% To incorporate .eps illustrations using PDFLaTeX, etc.
\usepackage[caption=false]{subfig}% Support for small, `sub' figures and tables
\usepackage[natbibapa,nodoi]{apacite}% Citation support using apacite.sty. Commands using natbib.sty MUST be deactivated first!
\renewcommand\bibfont{\fontsize{10}{12}\selectfont}% Bibliography support using natbib.sty
\makeatletter% @ becomes a letter
\def\NAT@def@citea{\def\@citea{\NAT@separator}}% Suppress spaces between citations using natbib.sty
\makeatother% @ becomes a symbol again
\usepackage{amsfonts}
\usepackage{amsthm}
\usepackage{amssymb}
\usepackage{setspace}

\theoremstyle{plain}% Theorem-like structures provided by amsthm.sty
\newtheorem{theorem}{Theorem}[section]
\newtheorem{lemma}[theorem]{Lemma}
\newtheorem{corollary}[theorem]{Corollary}

\theoremstyle{definition}

\theoremstyle{remark}
\newtheorem{remark}{Remark}

\AtBeginDocument{
	\let\oldref\ref
	\renewcommand{\ref}[1]{
		\IfBeginWith{#1}{fig:}%
		{{\color{blue}Figure~\oldref{#1}}}%
		\IfBeginWith{#1}{eqn:}{eq:}%
		{{\color{blue}\oldref{#1}}}%
		{\IfBeginWith{#1}{tab:}{{\color{blue}Table~\oldref{#1}}}{}}}%
}

\newcommand{\Ebld}{{\mbox{\bf E}}}
\newcommand{\var}{\mbox{\bf Var}}
\usepackage{color}

\newcommand{\argmin}{\operatornamewithlimits{\arg\min}}

\newcommand{\wh}{\widehat}
\usepackage{caption}

\usepackage{mathtools}
\usepackage{xcolor}
\usepackage{comment}
\usepackage{graphicx}
\usepackage{multirow}
\usepackage{adjustbox}
\usepackage[pagewise, displaymath, mathlines]{lineno}
\usepackage{subfig}
\usepackage{authblk}
\providecommand{\keywords}[1]{\textbf{\textit{Keywords:}} #1}
\newtheorem{corollary}[theorem]{Corollary}{\bfseries}{\itshape}
\newtheorem{lemma}[theorem]{Lemma}{\bfseries}{\itshape}
\newtheorem{remark}[theorem]{Remark}{\bfseries}{\itshape}

\usepackage[nottoc]{tocbibind}
\usepackage[a4paper,bindingoffset=0.2in,%
left=0.5in,right=0.5in,top=1in,bottom=1in,%
footskip=.25in]{geometry}
\usepackage{changepage}
\usepackage[nopar]{lipsum}
\usepackage{amsmath,amssymb,amsthm,enumitem}

\title{Estimating Linear Mixed Effects Models with Truncated Normally Distributed Random Effects}
\author[1]{Hao Chen\thanks{Corresponding Author: hao.x.chen@nielseniq.com}}
\author[1]{Lanshan Han\thanks{lanshan.han@nielseniq.com}}
\author[1,2]{Alvin Lim\thanks{aclim2@emory.edu}}
\affil[1]{Retail Product Research \& Development, NielsenIQ, Chicago, IL 60606}
\affil[2]{Emory University Goizueta Business School, Atlanta, GA 30322}
\date{} 
\begin{document}
\maketitle
\begin{abstract}
Linear Mixed Effects (LME) models have been widely applied in clustered data analysis in many areas including marketing research, clinical trials, and biomedical studies. Inference can be conducted using maximum likelihood approach if assuming Normal distributions on the random effects. However, in many applications of economy, business and medicine, it is often essential to impose constraints on the regression parameters after taking their real-world interpretations into account. Therefore, in this paper we extend the classical (unconstrained) LME models to allow for sign constraints on its overall coefficients. We propose to assume a symmetric doubly truncated Normal (SDTN) distribution on the random effects instead of the unconstrained Normal distribution which is often found in classical literature. With the aforementioned change, difficulty has dramatically increased as the exact distribution of the dependent variable becomes analytically intractable. We then develop likelihood-based approaches to estimate the unknown model parameters utilizing the approximation of its exact distribution. Simulation studies have shown that the proposed constrained model not only improves real-world interpretations of results, but also achieves satisfactory performance on model fits as compared to the existing model.
\end{abstract}

\noindent \keywords{Mixed Effects Model; Constrained Regression Analysis; Penalized Least Squares; Penalized Restricted Least Squares}

\section{Introduction} \label{sec:intro}
In practice, it is often necessary for modelers to quantify the heterogeneity among subgroups in a statistical study. For example, in marketing research, modelers often need to consider the geographical difference in response to marketing activities. In clinical research, a modeler may aim to capture the demographic difference in responding to a certain treatment regime. To do so, modelers often rely on mixed effect models with random effects specified to capture the heterogeneity. It is also often assumed that the random effects follow Normal distributions, so that a closed-form likelihood function is available and thus efficient numerical algorithms are available to maximize it. However, in many applications the normality assumption can lead to results that are inconsistent with our common sense or not interpretable. We provide an example to elaborate this issue.

We consider a study in marketing research with the goal of inferring the effectiveness of various marketing activities from sales and marketing data. Typically, for this kind of study, modelers collect weekly sales volume and marketing activity readings across different geographical regions for various products. A statistical model, typically a mixed effect model, is specified to find how the marketing activities affect the sales volume of the individual products in the different regions. We consider a simplified real-world dataset recording weekly sales and discount information on a consumer packaged good (CPG) \citep{bronnenberg2007consumer} from a retailer. The dataset contains weekly sales volume from different stores in three consecutive weeks of a non-holiday period in 2017. It includes three columns: (1) Store Cluster, (2) Discount Rate and (3) Logit Quantity. Each row represents a weekly record of a particular store. A snapshot of the dataset is given in Table~\oldref{tab:app2}. Some explanations about the variables are given below.
\begin{itemize}
	\item Store~Cluster: the stores are pre-clustered into $6$ different store clusters ($A, B, C, D, E, F$) based on their proximity such that stores within a cluster are more homogeneous than those  across cluster.
	\item Discount Rate: the discount rate ranging between $0$ and $1$. The larger the discount rate is, the cheaper the product is sold to customers. A $0$ rate means the product was not sold at a promotional discount for that week.
	\item Logit Quantity: the weekly sales quantity of a product after a logistic transformation, defined as $ \log\left( \frac{q}{\mu - q} \right) $, where $q$ is the observed sales quantity and $\mu$ is the theoretical maximum sales quantity from domain knowledge and treated as known.
\end{itemize}
\begin{table}[htbp!]
	\caption{Snapshot of the dataset with discounted sales information on a product.} \label{tab:app2}
	\centering
	\begin{tabular}{|c|c|c|} \hline \hline  
		Store Cluster & Discount Rate & Logit Quantity \\ \hline \hline 
		A             & 0.000 & 0.236               \\ \hline
		$\cdots$       & $\cdots$ & $\cdots$               \\ \hline
		A             & 0.000 & 0.358 \\ \hline \hline
		B          & 0.000 & 0.272               \\ \hline
		$\cdots$      & $\cdots$ & $\cdots$ \\ \hline
		B           & 0.440 & -0.333 \\ \hline \hline
		$\vdots$        & $\vdots$  & $\vdots$                \\ \hline \hline
		F          & 0.010 & 0.082               \\ \hline
		$\cdots$      & $\cdots$ & $\cdots$ \\ \hline
		F           & 0.000 & -1.113 \\ \hline \hline
	\end{tabular}
\end{table}
Given this dataset, we aim to quantify the relationship between Logit Quantity and Discount Rate while capturing heterogeneity across store clusters. Note that for simplicity of illustration, the dataset has been simplified with some other confounding factors removed to focus on the Discount Rate. The temporal effect is also ignored since neither seasonality nor holiday effects is expected to affect modeling results in a non-trivial way for the three-week non-holiday period, and is not the focus of this research. We specify a linear mixed effects (LME) model \citep{mcculloch2014generalized} with both random intercept and random slope in (\oldref{eq:mot}):
\begin{equation} \label{eq:mot}
	y_{\ell, i} = (\beta_0 + \beta_{0, \ell}) + (\beta_1 + \beta_{1, \ell}) x_{\ell, i} + \varepsilon_{\ell, i},
\end{equation}
where $ \ell $ is the index for Store Cluster, which is also the grouping factor, $x_{\ell, i}$ is the Discount Rate for observation $i$ and Store Cluster $\ell$, $ y_{\ell, i} $ is the Logit Quantity for observation $i$ and Store Cluster $\ell$ and $\varepsilon_{\ell, i}$ is the random error term. Working under the independent covariance structure \citep{wu2009mixed}, the classical LME model assumes $ \varepsilon_{\ell, i} \sim N(0, \sigma^2)$, $ \beta_{0, \ell} \sim N(0, \varsigma_0^2)$, $ \beta_{1, \ell} \sim N(0, \varsigma_1^2) $ and the random effects and the error term are independent to each other. 

In addition to the above assumptions, in practice, it is also required that $ \beta_{1} \ge 0 $ and the overall coefficient $  \beta_1 + \beta_{1, \ell} \ge 0$, representing a common belief that promotional discounts will not reduce sales quantity. In other words, a non-negative sign constraint is needed on the fixed effect and the overall coefficient of Discount Rate. A classical LME model was fitted on the dataset that produces $ \hat{\beta}_1 = -0.319$, and none of the absolute values of the estimated random effects is above $ 0.1 $ making all of the $ 6 $ overall coefficients negative as well. With the aforementioned modeling results, some heuristics often needs to be applied to ``correct" the sign of the estimated coefficients for Discount Rate before the model is considered as conceivable. This example demonstrates the necessity to deviate from the normality assumption when equipping the traditional LME models with sign constraints while ensuring the technical rigor of the statistical assumptions and estimation approaches.

The above motivating example is merely one of the many real-world applications, where one often has restrictions and/or prior knowledge on the signs of the parameters to be estimated when the business or physical interpretations are taken into account. In addition to practical benefits, a sign constraint also bear some theoretical merits. For example, it can help with model identifiability by truncating parts of the possible parameter space. It also mitigates the issue of multicollinearity by restricting the feasible regions. This is especially useful when data quality is a concern. 

The major contribution of this research is to estimate the model parameters with sign constraints assuming that the random effects follow a symmetric doubly truncated Normal (SDTN) distribution, instead of the unconstrained Normal distribution often found in the literature. The lower bounds and upper bounds of a SDTN distribution are carefully chosen based on its fixed effects so that the overall coefficients will not violate the sign constraints. The ``minor'' change in the  distribution on the random effects brings some profound differences in the estimation process, and has dramatically increased the difficulty as a SDTN distribution does not have as elegant a set of mathematical properties as the unconstrained Normal distribution does. As a simple illustrating example, the sum of two independent SDTNs is not necessarily a SDTN with known analytical expressions of its resulting parameters, for which people take it for granted in terms of an unconstrained Normal distribution. Therefore, it is practically infeasible to derive a closed form probability density function for the sum of several independent SDTNs. This is a major hurdle if a likelihood based estimation approach is to be applied. In this paper, we use an approximated probability density function inspired by the Central Limit Theorem (CLT) to derive an approximated likelihood function. Parameters estimation is then conducted by maximizing the approximated likelihood function.

The linear mixed effects model enjoys its popularity during the past several years, and it is not unexpected that there are quite many monographs and academic articles about it covering a full spectrum of areas from studies on its mathematical properties and computational aspects of its implementation to its applications in economics, marketing, medicine and pharmaceutical research. \cite{jiang2007linear} provided a comprehensive overview of the mixed effects model focusing on its mathematical properties. \cite{lindstrom1988newton} discussed the computational details and proposed the use of Newton-Raphson and EM algorithms. \cite{demidenko2013mixed} reviewed its many implementations in R, while \cite{bates2014fitting} specifically introduced the popular R package, \textit{lme4}. \cite{wu2009mixed} covered how the model behaves when missing data and measurement errors are present.  On the application side, \cite{brabec2008nonlinear} used it to study the natural gas consumption by individual customers. It was also applied by \cite{mitsumata2012effects} to research the effects of parental hypertension on longitudinal trends in offspring blood pressure. Moreover, linear mixed effects models have been extended to generalized linear mixed effects models, see \cite{mcculloch2014generalized}, to handle situations where the response variable is non-Normal, such as dummy (binomial) and count (Poisson). In addition, \cite{wolfinger1993generalized} discussed its estimation using a pseudo-likelihood approach. In this paper, we restrict our research to the linear mixed effects model as our current business applications do not require the generalized version. 

There are a few other studies on non-Normal random effect models.  \cite{pinheiro2001efficient} specifically discussed the multivariate t-distribution random effects. More recently, \cite{nelson2006use} applied the probability integral transformation (PIT) to estimate non-linear mixed effects models with non-Normal random effects. \cite{liu2008likelihood} then proposed a computationally more practical method to obtain the maximum likelihood estimations for mixed effects models by reformulating the conditional likelihood function on non-Normal random effects. Moreover, \cite{yucel2010impact} employed simulations to study the impact of random effects generated from non-Normal distributions on statistical inference under missing data conditions, while the Normality assumption is still assumed in the parameter estimation process. A direct comparison between the proposed methods and PIT is discussed further in Section~\oldref{sec:pit}.

Admittedly, it is often possible to impose sign constraints in the estimation process under the existing framework of the linear mixed effects model, namely, imposing the constraints without modifying the likelihood function in the unconstrained case. However, we argue that this is not a theoretical sound approach. In particular, under the classical framework, random effects are assumed to follow a Normal distribution. Imposing constraints on them requires us to deviate from the Normal distribution assumption and hence invalidate the widely used likelihood functions derived based on Normal distributions. Hence, we propose to use the SDTN distribution instead of the unconstrained Normal one on the random effects in the model specification to comply with the use of sign constraints in the estimation.

The rest of the paper is organized as follows. The model specification is given in Section~\oldref{sec:LHM}. We provide some theoretical results on the truncated Normal distribution in Section~\oldref{sec:TND}, building a solid foundation. The proposed estimation methods are detailed in Section~\oldref{sec:methods}. Some simulation results and application examples are presented in Section~\oldref{sec:simulation} and Section~\oldref{sec:app}, respectively. We make several concluding remarks in Section~\oldref{sec:conclusions}.

\section{Linear Mixed Effects Model with Sign Constraints}\label{sec:LHM}
Let $\mathbf 0_n$ be the all zero vector of length $n$, $\mathbf 1_n$ be the all one vector of length $n$, $\mathbf I_n$ be the identity matrix of size $n\times n$. Throughout this paper, we use lower case letters to represent scalars, bold lower case letters to represent vectors, and upper case letters to represent matrices. For a vector $\boldsymbol x\in \mathbb R^n$, let $\mbox{diag}[\boldsymbol x]$ be the $n\times n$ diagonal matrix with the main diagonal being $\boldsymbol x$. We use subscripts to index scalars and superscripts to index vectors and matrices. For a vector $\boldsymbol x$, $x_i$ represents its $i$-th component. For a matrix $X \in \mathbb R^{n\times p}$, $x_{i,j}$ represents its $(i,j)$ element. Let $\alpha \subseteq \{1,\cdots, p\}$ be an index set and $\overline \alpha$ be its complement. We write as $x_\alpha$ the vector formed by taking the elements with indices in $\alpha$ from $x \in \mathbb R^p$  , and $X_{\bullet, \alpha}$ the submatrix of $X$ composed of the columns with indexes in $\alpha$ from $X \in \mathbb R^{n \times p}$. 

Consider a classical linear mixed effect model that captures heterogeneity among different clusters. Let the clusters, which could be geographical regions, product classes, individual subjects, etc, be indexed by $\ell=1,\cdots,g$. For each cluster $\ell$, the dependent variable $\boldsymbol{y}^\ell$ is linearly dependent on the independent variables with an error term following a Normal distribution with $0$ mean and unknown variance (to be estimated). Mathematically, this model is given by:
\begin{equation}\label{eq:mixed_reformulation_matrix2}
\boldsymbol y^\ell \, =\, X^\ell \boldsymbol \beta + Z^\ell \boldsymbol \gamma^\ell + \boldsymbol \varepsilon^\ell,
\end{equation}
where
\begin{equation}\nonumber
\boldsymbol y^\ell \,\triangleq \, \left[
\begin{array}{c}
y_{\ell,1}\\
\vdots\\
y_{\ell,n_\ell}
\end{array}
\right] \, \in \, \mathbb R^{n_\ell}, \,\,
\boldsymbol \beta \, \triangleq \, \left[
\begin{array}{c}
\beta_{1}\\
\vdots\\
\beta_{p}
\end{array}
\right] \, \in \, \mathbb R^{p}, \,\,
\boldsymbol \gamma^\ell \, \triangleq \, \left[
\begin{array}{c}
\gamma_{\ell,1}\\
\vdots\\
\gamma_{\ell,k}
\end{array}
\right] \, \in \, \mathbb R^{k},
\end{equation}
\begin{equation}\nonumber
X^\ell \, \triangleq \,
\left[
\begin{array}{cccc}
x_{\ell, 1, 1} &  x_{\ell, 1, 2}  &  \cdots  &  x_{\ell, 1, p} \\
\vdots  &   \vdots  &  \vdots  &  \vdots \\
x_{\ell, n_\ell, 1} &  x_{\ell, n_\ell, 2}  &  \cdots  &  x_{\ell, n_\ell, p}
\end{array}
\right] \, \in \, \mathbb R^{n_\ell \times p},
\end{equation}
\begin{equation}\nonumber
Z^\ell \, \triangleq \,
\left[
\begin{array}{cccc}
z_{\ell, 1, 1} &  z_{\ell, 1, 2}  &  \cdots  &  z_{\ell, 1, k} \\
\vdots  &   \vdots  &  \vdots  &  \vdots \\
z_{\ell, n_\ell, 1} &  z_{\ell, n_\ell, 2}  &  \cdots  &  z_{\ell, n_\ell, k}
\end{array}
\right] \, \in \, \mathbb R^{n_\ell \times k},
\end{equation}
\begin{equation}\nonumber
\mbox{and} \,\,
\boldsymbol{\epsilon}^\ell \, \triangleq \,
\left[
\begin{array}{c}
\epsilon_{\ell, 1}  \\
\vdots  \\
\epsilon_{\ell, n_\ell}
\end{array}
\right] \, \in \, \mathbb R^{n_\ell}.
\end{equation}
We often call $ \boldsymbol{\beta}$ the fixed effects, $\gamma_{\ell, i}, \ell = 1, \ldots, g; i = 1, \ldots, k $ the random effects, and
$ \boldsymbol{\epsilon}^{\ell} \sim \mathcal N( \mathbf 0_{n_\ell}, \sigma^2 \mathbf I_{n_\ell})$ the error vector with $\sigma^2$ unknown. Note that $ n_{\ell} $ is the sample size for group $ \ell $, and the total size is $ n = \sum_{\ell=1}^{g} n_{\ell} $. $ p $ is the dimension. The number of variables for which random effects will be considered is denoted by $ k, 0 \le k \le p $. If $ k=0 $, the linear mixed effects model reduces to a linear regression model with fixed effects only. One could also include an intercept to measure the baseline. The classical linear mixed effects model also assumes that
\begin{equation}
\boldsymbol{\gamma}^{\ell} \overset{iid}{\thicksim} \mathcal N( \mathbf{0}_k, \Sigma), \,\, \forall \, \ell=1,\cdots,g,
\end{equation}
where $ \Sigma$ is parameterized by some unknown parameters. Stacking up the data from different clusters, we have
\begin{equation}\nonumber
\boldsymbol y \,\triangleq \, \left[
\begin{array}{c}
\boldsymbol y^1\\
\vdots\\
\boldsymbol y^g
\end{array}
\right]  \in  \mathbb R^{n}, \,\,
\boldsymbol \gamma \, \triangleq \, \left[
\begin{array}{c}
\boldsymbol \gamma^1\\
\vdots\\
\boldsymbol \gamma^g
\end{array}
\right]  \in  \mathbb R^{kg},
\,\,
X \, \triangleq \, \left[
\begin{array}{c}
X^1\\
\vdots\\
X^g
\end{array}
\right]  \in  \mathbb R^{n\times p},
\end{equation}
\begin{equation}\nonumber
\mbox{and} \,
Z \, \triangleq \,
\left[
\begin{array}{ccc}
Z^1 &  \\
&   \ddots  &  \\
&&  Z^g
\end{array}
\right] \, \in \, \mathbb R^{n \times kg},
\boldsymbol \epsilon \,\triangleq \, \left[
\begin{array}{c}
\boldsymbol \epsilon^1\\
\vdots\\
\boldsymbol \epsilon^g
\end{array}
\right]  \in  \mathbb R^{n}, \,\,
\end{equation}
With these definitions, we can rewrite the model into a more concise expression given below.
\begin{equation}\label{eq:LMM}
\boldsymbol{y} \, = \, X\boldsymbol \beta + Z\boldsymbol \gamma + \boldsymbol \epsilon,
\end{equation}
where $ \boldsymbol{\gamma} \sim \mathcal N( \mathbf{0}_{n}, G), G \, = \,
\left[
\begin{array}{ccc}
\Sigma \\
&\ddots \\
&& \Sigma
\end{array}
\right]$ and
$  \boldsymbol \epsilon \sim \mathcal N(\mathbf{0}_n, R)  $ with $  R = \sigma^2 \mathbf{I}_{n}$. In addition, $\boldsymbol \gamma$ and $\boldsymbol \varepsilon$ are independent, and hence
\begin{equation} \nonumber 
\left( \begin{array}{c}
\boldsymbol \gamma\\
\boldsymbol \epsilon
\end{array}\right) \, \sim \, \mathcal N
\left(0,
\left[\begin{array}{cc}
G & 0\\
0 & R
\end{array}\right]\right).
\end{equation}

\subsection{\textbf{Maximum Likelihood Approach for Linear Mixed Effects Models: a Brief Review}}
Working under the model specification in (\oldref{eq:LMM}), a classical approach to estimate the parameters is to maximize certain likelihood functions. According to the assumptions, $\boldsymbol{y}$ also follows a multivariate Normal distribution, in particular,
\begin{equation} \nonumber 
\boldsymbol y \, \sim \, \mathcal N(X\boldsymbol \beta, ZGZ^T+R),
\end{equation}
where $ G $ is parameterized by the unknown $ \varsigma_1^2, \ldots, \varsigma_k^2 $ and $ R $ is parameterized by $ \sigma^2 $. Let $ \boldsymbol{\theta} $ collectively denote $ \varsigma_1^2, \ldots, \varsigma_k^2 $ and $ \sigma^2 $. The covariance matrix of $\boldsymbol y$ can hence be written as
\begin{equation} \nonumber
V(\boldsymbol \theta) \, \triangleq \, ZG(\boldsymbol \theta)Z^T+R(\boldsymbol \theta).
\end{equation}
For any squared matrix $A$, let $|A|$ be its determinant. The \textbf{profile} log-likelihood function is defined by:
\begin{equation}\label{eq:likelihood}
\mathcal L(\boldsymbol \theta | \boldsymbol{y}, \boldsymbol{X}, \boldsymbol{Z} ) \, \triangleq \,  -\frac{1}{2}\left(\ln|V(\boldsymbol \theta)| + (\boldsymbol y-X\widetilde{\boldsymbol\beta} (\boldsymbol \theta))^TV(\theta)^{-1}(\boldsymbol y-X\widetilde {\boldsymbol\beta}(\boldsymbol \theta))\right),
\end{equation}
where
\begin{equation}\label{eq:beta_of_theta}
\widetilde{\boldsymbol \beta}(\boldsymbol \theta) = (X^TV(\boldsymbol \theta)^{-1}X)^{-1}X^TV(\boldsymbol \theta)^{-1} \boldsymbol y.
\end{equation}
According to \cite{zhang2015tutorial}, the \textbf{restricted} log-likelihood function is defined
\begin{equation}\label{eq:RE_likelihood}
\mathcal L_{\mbox{REML}}(\boldsymbol \theta | \boldsymbol{y}, \boldsymbol{X}, \boldsymbol{Z}) \triangleq
-\frac{1}{2}\left(\ln|V(\boldsymbol \theta)| + (\boldsymbol y-X\widetilde{\boldsymbol\beta} (\boldsymbol \theta))^TV(\boldsymbol \theta)^{-1}(\boldsymbol y-X\widetilde{\boldsymbol \beta}(\boldsymbol \theta))+ \ln|X^TV(\boldsymbol \theta)^{-1}X|\right).
\end{equation}
By maximizing $\mathcal L(\boldsymbol \theta)$ or $\mathcal L_{\mbox{REML}}(\boldsymbol \theta)$ using either Newton's method \citep{lindstrom1988newton,wolfinger1994computing} or EM algorithm \citep{dempster1977maximum}, we can obtain the estimator of $\boldsymbol \theta$, i.e.,
\begin{equation} \nonumber
\widehat{\boldsymbol\theta}_{\mbox{ML}} = \mbox{argmax} \, \mathcal L(\boldsymbol \theta), \, \mbox{ or }\,\, \widehat {\boldsymbol\theta}_{\mbox{REML}} = \mbox{argmax} \, \mathcal L_{\mbox{REML}}(\boldsymbol \theta).
\end{equation}
Based on \cite{mcculloch2014generalized}, it is known that $\widehat{\boldsymbol\theta}_{\mbox{ML}}$ is \textbf{biased}, while $\widehat{\boldsymbol\theta}_{\mbox{REML}}$ is \textbf{unbiased} which is more favorable in theory. After estimates of $\boldsymbol \theta$ (denoted $\widehat{ \boldsymbol \theta}$) are obtained, we can let $\widehat G= G(\widehat{\boldsymbol\theta})$ and $\widehat R= R(\widehat{\boldsymbol \theta}) $ and estimate $\boldsymbol\beta$ and $\boldsymbol\gamma$ together by maximizing the joint likelihood function with $\widehat G$ and $\widehat R$ considered as known, i.e.,
\begin{equation}
\begin{array}{l}
f(\boldsymbol y,\boldsymbol \gamma) \, = \, f_{\boldsymbol y}(\boldsymbol y|\boldsymbol \gamma)f_{\boldsymbol \gamma}(\boldsymbol \gamma) \nonumber \\[5pt]
\,\propto \, \displaystyle{ |\widehat G|^{-\frac{1}{2}}\exp\left(-\frac{1}{2} \boldsymbol\gamma^T \widehat G^{-1} \boldsymbol \gamma\right) |\widehat R|^{-\frac{1}{2}} \exp \left(-\frac{1}{2} (\boldsymbol y-X\boldsymbol \beta-Z\boldsymbol \gamma)^T \widehat R^{-1}(\boldsymbol y-X\boldsymbol \beta-Z\boldsymbol \gamma)\right) }. \nonumber
\end{array}
\end{equation}
Ignoring constant terms, the joint log likelihood function of $ \boldsymbol \beta$ and $\boldsymbol \gamma $ is hence,
\begin{equation} \nonumber
\ln f(\boldsymbol y,\boldsymbol \gamma) \propto  -\frac{1}{2} \boldsymbol \gamma^T \widehat G^{-1} \boldsymbol \gamma \, - \, \frac{1}{2}(\boldsymbol y-X\boldsymbol \beta-Z\boldsymbol \gamma)^T \widehat R^{-1}(\boldsymbol y-X\boldsymbol \beta-Z\boldsymbol \gamma). 
\end{equation}
Therefore, it suffices to minimize the following function with respect to $\boldsymbol \beta$ and $\boldsymbol \gamma$. 
\begin{equation}
Q(\boldsymbol \beta,\boldsymbol \gamma) =  \boldsymbol \gamma^T \widehat G^{-1} \boldsymbol \gamma \, + \, (\boldsymbol y-X\boldsymbol \beta-Z\boldsymbol \gamma)^T \widehat R^{-1}(\boldsymbol y-X\boldsymbol \beta-Z\boldsymbol \gamma). \nonumber
\end{equation}
By taking the first-order derivatives of $Q$ with respect to $\boldsymbol{\beta}$ and $\boldsymbol{\gamma}$, and set them equal to zero, we obtain the following linear system:
\begin{equation}\label{eq:joint_beta_gamma_system}
\left[
\begin{array}{cc}
X^T\widehat R^{-1} X & X^T\widehat R^{-1} Z \\
Z^T \widehat R^{-1} X & Z^T \widehat R^{-1} Z + \widehat G^{-1}
\end{array}
\right] \left[
\begin{array}{c}
\boldsymbol\beta\\
\boldsymbol\gamma
\end{array}
\right] \, = \, \left[
\begin{array}{c}
X^T\widehat R^{-1} y\\
Z^T \widehat R^{-1} y
\end{array}
\right]
\end{equation}
Let $\widehat V = V(\widehat \theta)$, we can derive from (\oldref{eq:joint_beta_gamma_system}) that the estimators of $\boldsymbol{\beta}$ and $\boldsymbol{\gamma}$ as follows.
\begin{eqnarray}
\widehat{\boldsymbol \beta} & = & (X^T \widehat V^{-1}X)^{-1}X^T \widehat V^{-1}\boldsymbol y, \label{eq:beta_estimation}\\
\widehat{ \boldsymbol\gamma} & = & \widehat GZ^T \widehat V^{-1}(\boldsymbol y-X\widehat{\boldsymbol \beta}). \label{eq:gamma_estimation}
\end{eqnarray}
The aforementioned classical approach has been successfully used in many applications, and has been implemented in a statistical software packages such as R, Python and SAS. However, as discussed in Section \oldref{sec:intro}, practitioners often face the situation where sign constraints are needed for some of the unknown coefficients to make practical sense. Hence, in what follows, we will augment the approach to handle sign constraints in a mathematical rigorous way.

\subsection{\textbf{Proposed Model Specification}}
In this paper, we consider a specific case where $Z^\ell$ is usually formed by taking a subset of the columns from $ {X}^{\ell} $, i.e. $ {Z}^{\ell} = X^\ell_{\bullet, \alpha} $ for some index set $\alpha \subseteq \{1,\cdots,p\}$. Mixed effects are considered for any column indexed with $ \alpha $. In this case, for each $\ell = 1,\cdots, g$, model (\oldref{eq:mixed_reformulation_matrix2}) can be written as
\begin{equation} \nonumber
\boldsymbol y^\ell \, = \, X^\ell_{\bullet, \alpha} (\beta_{\alpha} + \gamma_{\alpha}^{\ell}) + X^\ell_{\bullet, \overline{\alpha}} \beta_{\overline{\alpha}} + \boldsymbol \varepsilon^\ell.
\end{equation}
In many applications, it is necessary to make sure that the elements in $\beta_\alpha + \gamma_\alpha^\ell$ have correct signs. Without loss of generality, we consider the situation where $\alpha =\{1,\cdots, p\}$ and $\beta_\alpha+\gamma_\alpha^\ell \geq 0$. In other words, a full model in which mixed effects are accounted for every column is assumed. The subscript $ \alpha $ is dropped for the remainder of this section. We further assume
\begin{equation} \nonumber
\Sigma  = \left[\begin{array}{ccc} \varsigma_1^2 \\ & \ddots \\&& \varsigma_p^2 \end{array}\right],
\end{equation}
where $ \varsigma_i^2  $'s are unknown. We continue to follow the classical model by assuming that 
\begin{equation}\label{eq:mixed_reformulation_matrix_proposed}
\boldsymbol y^\ell \, =\, X^\ell \boldsymbol \beta + Z^\ell \boldsymbol \gamma^\ell + \boldsymbol \varepsilon^\ell,
\end{equation}
where
\begin{equation}
\boldsymbol \varepsilon^\ell \sim \mathcal{N}( \mathbf{0}_{n_\ell}, \sigma^2 \mathbf{I}_{n_\ell}).
\end{equation}
However, to ensure the nonnegativity requirment is satisfied, we assume that the random effects $ \gamma_{\ell, i} $ independently follow a symmetric doubly truncated normal (SDTN) distribution that is bounded by $ -|\beta_i|$ and $ |\beta_i| $. We will explicitly define SDTN in Section \oldref{sec:TND}. As a result, each $ \gamma^{\ell}_i $ is mathematically constrained within its corresponding $[-|\beta_i|, |\beta_i| ]$. This way, we can guarantee that the overall coefficient will be non-negative. The proposed modification seems to be rather straightforward, however, it imposes significant challenges in the estimation of the parameters. Specifically, the distribution of the sum of finitely many random variables following SDTN distributions does not have a concise closed form. This fact represents a major hurdle when a maximum likelihood approach is applied. We will elaborate on more details about the parameter estimation process in Section~\oldref{sec:methods}.

We will study and present results on some fundamental properties regarding truncated Normal distribution in Section~\oldref{sec:TND} to lay a solid foundation, upon which our proposed estimating methods will build.

\section{Truncated Normal Distributions}\label{sec:TND}
In this section, we present some properties of the truncated Normal distribution of interests. Throughout the paper, whenever we use the term Normal distribution, we refer to the usual unconstrained Normal distribution unless specified otherwise. Let the error function for a Normal distribution be
\begin{equation} \nonumber
\mbox{erf}(z) \, \triangleq \, \frac{1}{\sqrt{\pi}} \int_{0}^{z} e^{-t^2} dt.
\end{equation}
A truncated Normal (TN) distribution is parameterized by $ 4 $ parameters: location, $ \mu $; scale, $ \eta $; lower bound $ a $; upper bound $ b $. The Normal distribution is a special case of the TN distribution with $ a = -\infty  $ and $  b = \infty $. The probability density function (PDF) of a $\mathcal {TN}(\mu, \eta^2, [a,b])$, with $\eta>0$, is given by
\begin{equation} \nonumber
f_{\mathcal {TN}}(x; \mu,\eta^2, a, b) \, = \, \left\{
\begin{array}{ll}
\displaystyle{ \frac{1}{\eta} \frac{\phi\left(\xi\right)}{\Phi\left(b^{\prime}\right)-\Phi\left(a^{\prime}\right)} }, & x\in[a,b]\\[5pt]
0, & \mbox{otherwise}
\end{array}\right.,
\end{equation}
where $\phi(\cdot)$ and $\Phi(\cdot)$ are the PDF and cumulative distribution function (CDF) of the standard Normal distribution, i.e.,
\begin{equation} \nonumber
\phi(\xi) \, = \, \frac{1}{\sqrt{2\pi}} \exp\left(-\frac{1}{2}\xi^2\right), \,\, \mbox{and} \,\, \Phi(\xi) \, = \, \frac{1}{2} \left[1+\mbox{erf}\left(\frac{\xi}{\sqrt 2}\right)\right],
\end{equation}
respectively, and
\begin{equation} \nonumber
\xi \, \triangleq \, \frac{x-\mu}{\eta}, \, \, a^{\prime} \, \triangleq \, \frac{a-\mu}{\eta},\, \, \mbox{ and } \,\,b^{\prime} \, \triangleq \, \frac{b-\mu}{\eta}.
\end{equation}
The mean and variance of $x \sim \mathcal{TN}(\mu, \eta^2, [a,b])$ are known and given by \cite{olive2008applied}:
\begin{equation} \nonumber
\Ebld[x] \, = \, \mu + \frac{\phi(a^{\prime}) - \phi(b^{\prime})}{\Phi(b^{\prime}) - \Phi(a^{\prime})} \eta,
\end{equation}
\begin{equation} \nonumber
\var[x] \, = \, \eta^2 \left[ 1+ \frac{a^{\prime} \phi(a^{\prime}) - b^{\prime} \phi(b^{\prime})}{\Phi(b^{\prime}) - \Phi(a^{\prime})} -\left(\frac{\phi(a^{\prime})-\phi(b^{\prime})}{\Phi(b^{\prime}) - \Phi(a^{\prime})}\right)^2\right].
\end{equation}
In this paper, we are particularly interested in a special case, namely the symmetric doubly truncated normal (SDTN) distribution $\mathcal {TN} (\mu, \eta^2, [\mu - \rho \eta, \mu+\rho \eta] )$ with $\rho>0$, denoted by $\mathcal {SDTN}(\mu, \eta^2, \rho)$. It is a special case of a TN with $ a = \mu - \rho \eta, b = \mu + \rho \eta $, i.e., the lower bound and upper bound are symmetric around mean $ \mu $. The properties of a SDTN distrbution is given by Lemma \oldref{lm:SDTN_properties1}.
\begin{lemma}\label{lm:SDTN_properties1}
	Suppose $x \sim \mathcal{SDTN}(\mu, \eta^2, \rho)$, the following results hold
	\begin{itemize}
		\item[(i).] The density function is 
		\begin{equation} \nonumber
		f_{\mathcal{SDTN}}(x;\mu, \eta^2, \rho) \,= \, \left\{ \begin{array}{ll}\displaystyle{\frac{1}{\eta} \frac{\phi(\xi)}{2\Phi(\rho)-1} }, & x\in[\mu-\rho \eta, \mu+\rho \eta]\\
		0, & \mbox{otherwise} \end{array}\right\}
		\end{equation}
		\item[(ii).] The expectation is  
		\begin{equation} \nonumber
		\Ebld[x] \, = \, \mu,
		\end{equation}
		\item[(iii).] The variance is 
		\begin{equation} \nonumber
		\var[x] \, = \, \eta^2 \left[1 - \frac{2\rho \phi(\rho)}{2\Phi(\rho)-1}\right],
		\end{equation}
	\end{itemize}
\end{lemma}
The proof is omitted as it is straightforward to verify the above results. Note that we define SDTN distributions with $\rho>0$. In fact, when $\rho=0$, it becomes a deterministic value, and hence the variance is $0$. This is indeed consistent with the fact that
\begin{equation} \nonumber
\lim_{\rho\rightarrow 0} \left[ 1 - \frac{2\rho \phi(\rho)}{2\Phi(\rho)-1} \right] \, = \, 1 - \lim_{\rho \rightarrow 0 }\frac{2\rho \phi(\rho)}{2\Phi(\rho)-1} \, = \, 1- \lim_{\rho \rightarrow 0} \frac{2\phi(\rho) + 2\rho \phi'(\rho)}{2\phi(\rho)} \, = \, 0
\end{equation}
where the second equal sign is due to L'H$\hat{o}$pital's rule.
We state some properties regarding the SDTN distribution in Lemma \oldref{lm:SDTN_properties2}.
\begin{lemma}\label{lm:SDTN_properties2}
	Let $x \sim \mathcal{SDTN}(\mu, \eta^2, \rho)$ with $\rho>0, \eta > 0$, the following properties hold:
	\begin{itemize}
		\item[(i).] $x-\mu \sim \mathcal{SDTN}(0, \eta^2, \rho)$.
		\item[(ii).] For any $x, y \in[\mu-\rho\eta, \mu+\rho \eta]$, if $x+y = 2 \mu$ then $f_{\mathcal{SDTN}}(x;\mu, \eta^2, \rho) = f_{\mathcal{SDTN}}(y;\mu, \eta^2, \rho)$.
		\item[(iii).] $\var[x] \leq \eta^2$.
		\item[(iv).] Suppose $x' \sim \mathcal{SDTN}(\mu, \eta^2, \rho')$, then $\var[x] \leq \var[x']$ if $\rho \leq \rho'$.
		\item [(v).] If $x \sim \mathcal{SDTN}(0, \eta^2, \rho)$. Define $ x^{\prime} = k_0 + k_1 x $ with $ k_1 \ne 0$, where $ k_0, k_1 $ are finite real numbers, then $  x^{\prime} \sim \mathcal{SDTN}(k_0, k_1^2\eta^2, \rho) $.
	\end{itemize}
\end{lemma}
The proof of Lemma \oldref{lm:SDTN_properties2} is provided in Appendix $A$ of the \textit{Supplemental Document}. It is worth pointing out that, while the sum of independent non-identically distributed Normal random variables is Normally distributed, it is not the case for SDTNs. The exact distribution of the sum of independent non-identically SDTNs is analytically intractable. However, the following Normality results hold.
\begin{theorem} \label{th:clt}
	For every $ x_i \sim \mathcal{SDTN}(\mu_i, \eta_i^2, \rho_i)$, the random variables making up the collection $  \mathbf{X}_n = \{ x_i:  1 \le i \le n \} $ are independent with the following conditions:
	\begin{itemize}
		\item $ \mu_i $ are finite, i.e., $ \max_{1 \le i \le n} \mu_i < +\infty$
		\item $ \rho_i $ is bounded from below by $ \underline{\rho} > 0 $
		\item $ \eta_i $ is bounded from below and above by $ \underline{\eta} > 0 $ and $ \bar{\eta} < + \infty $, respectively.
	\end{itemize}
	Then
	\begin{equation} \nonumber
	\frac{1}{t_n} \sum_{i=1}^{n} \left(x_i - \mu_i\right) \, \xrightarrow{d} \, \mathcal N(0, 1),
	\end{equation}
	as $n\rightarrow \infty$, where $$t_n^2 = \sum_{i=1}^n \var[x_i].$$
\end{theorem}
The proof of Theorem \oldref{th:clt} is provided in Appendix $B$ of the \textit{Supplemental Document}. Moreover, it is also straightforward to verify the following corollary to Theorem~\oldref{th:clt}.
\begin{corollary}\label{co:weighted_CLT}
	Let $x_i \sim \mathcal{SDTN}(\mu_i, \eta_i^2, \rho_i)$, $i=1,2,\cdots$ be independent with $\mu_i$'s, $\eta_i$'s, and $\rho_i$'s satisfying conditions in Theorem \oldref{th:clt}. Let $\beta_i, i=1,2,\cdots,$ be real numbers and the absolute values are bounded from below and above, i.e., there exist $\overline \beta$ and $\underline \beta$ satisfying $ 0 < \underline{\beta} \leq |\beta_i| \leq \overline{\beta} < +\infty$ for all $i=1,2,\cdots$. Then,
	\begin{equation}
	\frac{1}{t_n} \sum_{i=1}^{n} \beta_i (x_i - \mu_i) \, \xrightarrow{d} \, \mathcal N(0, 1),
	\end{equation}
	as $n\rightarrow \infty$,  where 
	\begin{equation} \nonumber
	t_n^2 = \sum_{i=1}^n \beta_i^2 \var[x_i]. \qed
	\end{equation} 
\end{corollary}
Corollary~\oldref{co:weighted_CLT} indicates that the (weighted) sum of finitely many independent but non-identically distributed SDTNs converges in distribution to a Normal distribution. \\

Theorem \oldref{th:clt} and Corollary \oldref{co:weighted_CLT} indicate that the (weighted) sum of finitely many random variables obeying SDTN distributions, while does not exactly follow a Normal distribution, approximately obeys a Normal distribution when the number of the random variables in the summation is large. For conciseness, let's only focus on one row of the data: in model (\oldref{eq:LMM}), the response $y$ is approximately Normal given $\beta_i$'s. It is easy to see that
\begin{equation} \nonumber
\Ebld[\gamma_i] \, = \, 0,\,\, \mbox{ and } \,\, \var[\gamma_{i}]\, =\, \varsigma_i^2 \left[1-\frac{2\rho_i \phi(\rho_i)}{2\Phi(\rho_i)-1}\right],
\end{equation}
where $\rho_i = \frac{\beta_i}{\varsigma_i}, i = 1, \ldots, k$. Therefore, the mean and variance of $y$ given $\beta_i$'s and $x_i$'s are as follows
\begin{eqnarray}
\Ebld[y|\beta_i,x_i] & = & \sum_{i=1}^p \beta_i x_i, \nonumber \\
\var[y|\beta_i,x_i] & = & \sum_{i=1}^k x_i^2 \var[\gamma_{\ell,i}] + \var[\epsilon] \, = \, \sigma^2 + \sum_{i=1}^k x_i^2 \varsigma_i^2  \left[1-\frac{2\rho_i \phi(\rho_i)}{2\Phi(\rho_i)-1}\right]. \nonumber
\end{eqnarray}
With the observed data, if we write model in a matrix format, we have
\begin{eqnarray}
\Ebld[y|X, \boldsymbol{\beta}] & = & X \boldsymbol \beta, \nonumber \\
\var[y|X, Z, \boldsymbol{\beta}] & = & Z \Lambda Z^T + \sigma^2 \mathbf I_{n},\nonumber
\end{eqnarray}
where 
\begin{equation} \nonumber
\Lambda = \left[
\begin{array}{ccc}
\Delta \\
& \ddots \\&& \Delta
\end{array}\right]\mbox{ and } \Delta \, = \, \mbox{diag}\left[\left(\varsigma_i^2  \left[1-\frac{2\rho_i \phi(\rho_i)}{2\Phi(\rho_i)-1}\right]\right)_{i=1}^k\right].
\end{equation}
We approximate the distribution of $y$ given data $X, Z$ and all model parameters by a multivariate Normal distribution $\mathcal N(X\boldsymbol \beta, Z \Lambda Z^T + \sigma^2 \mathbf I_{n})$. With this approximation, we will report the proposed estimation methods in the next Section.

\section{Proposed Estimation Methods} \label{sec:methods}
We let $\boldsymbol \varsigma = (\varsigma_i)_{i=1}^k \in \mathbb R^k$. With an approximated distribution of $\boldsymbol y$ given the data $(X,Z)$ and the parameters, $\mathcal N(X\boldsymbol\beta, Z\Lambda Z^T + \sigma^2 \mathbf I_n)$, we estimate the unknown parameters by maximizing the approximated log-likelihood function. In fact, the approximated log-likelihood function is given by:
\begin{equation}\label{eq:approx_log_likelihood}
\mathcal L_{\mbox{approx}} (\boldsymbol \beta, \boldsymbol \varsigma, \sigma)\, = \, -\frac{n}{2} \ln 2\pi - \frac{1}{2}\ln|V| - \frac{1}{2}(\boldsymbol y-X\boldsymbol\beta)^TV^{-1}(\boldsymbol y-X\boldsymbol\beta),
\end{equation}
with $V \triangleq Z\Lambda Z^T + \sigma^2 \mathbf I_n$. Due to the requirement that $\boldsymbol \beta \geq 0$, it is necessary to keep $\boldsymbol \beta$ in the likelihood function explicitly instead of profiling it out as in the unconstrained case, i.e., Equation (\oldref{eq:likelihood}). Therefore, we propose to obtain estimates of $\boldsymbol \beta$, $\boldsymbol \varsigma$, and $\sigma$ by solving the following constrained optimization problem:
\begin{equation} \label{eqn:pls}
\begin{array}{rl}
(\widehat{\boldsymbol\beta}, \widehat{\boldsymbol \varsigma}, \widehat{\sigma}) \, = \, \displaystyle { \argmin_{\boldsymbol{\beta}, \boldsymbol{\varsigma}, \sigma} }  &
(\boldsymbol{y} - X\boldsymbol{\beta})^TV^{-1}(\boldsymbol{y} - X\boldsymbol{\beta})+ \ln|V|\\[5pt]
\mbox{s.t.}  & \boldsymbol \beta \, \geq \, 0,\\
&\boldsymbol \varsigma \, \geq \, 0,
\end{array}
\end{equation}
with $V \triangleq Z\Lambda Z^T + \sigma^2 \mathbf I_n$. Note that there is no need to impose nonnegativity on $\sigma$ in (\oldref{eqn:pls}) because only $\sigma^2$ appears in the objective function. However, it is not the case for $\boldsymbol\varsigma$ due to the definition of $\rho_i$'s. We refer to this approach as the penalized least squares (PLS) since (\oldref{eqn:pls}) can also be interpreted as minimizing least squares while penalizing on large $ \Lambda $ and $ \sigma^2 $. In addition, inspired by the restricted log-likelihood function for the unconstrained case, We propose as a second approach to solve the following constrained optimization problem.
\begin{equation} \label{eqn:reML}
\begin{array}{rl}
(\widehat{\boldsymbol\beta}, \widehat{\boldsymbol \varsigma}, \widehat{\sigma}) \, = \, \displaystyle { \argmin_{\boldsymbol{\beta}, \boldsymbol{\varsigma}, \sigma} }  &
(\boldsymbol{y} - X\boldsymbol{\beta})^TV^{-1}(\boldsymbol{y} - X\boldsymbol{\beta})+ \ln|V| + \ln|X^TV^{-1}X|\\[5pt]
\mbox{s.t.}  & \boldsymbol \beta \, \geq \, 0,\\
&\boldsymbol \varsigma \, \geq \, 0.
\end{array}
\end{equation}
In a similar interpretation, we refer to the second approach as penalized restricted least squares (PRLS). Let the optimal solution, most likely only local optimum due to the non-convexity of the objective functions, of either (\oldref{eqn:pls}) or (\oldref{eqn:reML}) be denoted as $(\widehat{\boldsymbol \beta},\widehat{\boldsymbol \varsigma},\widehat{\sigma})$, we can then estimate the random effect coefficients $\boldsymbol \gamma$. In fact, the joint likelihood function in this case is
\begin{equation}\nonumber
\prod_{\ell=1}^{g} \prod_{j=1}^{n_\ell} \left[\frac{1}{\widehat{\sigma} \sqrt{2\pi}} \exp\left(-\frac{1}{2}\left(\frac{y_{\ell,j}-\sum_{i=1}^p \left(\wh \beta_i + \gamma_{\ell,i}\right)x_{\ell,j,i} }{\widehat\sigma}\right)^2\right) \prod_{i=1}^p \frac{1}{\widehat\varsigma_i (2 \Phi(\widehat\rho_i)-1)} \phi\left(\frac{\gamma_{\ell,i}}{\widehat \varsigma_i}\right)\right],
\end{equation}
where $\wh \rho_i \triangleq \frac{\wh \beta_i}{\wh \varsigma_i}$. Therefore the joint log-likelihood function after removing constants (with respect to $\boldsymbol \gamma$) is given by:
\begin{equation}\label{eq:likelihood_gamma}
\mathcal L_{\boldsymbol \gamma}(\boldsymbol \gamma) \, = \, \sum_{\ell=1}^g \sum_{j=1}^{n_\ell}-\frac{1}{2}\left[\left( \frac{\widetilde y_{\ell,i} - \sum_{i=1}^{p} \gamma_{\ell,i} x_{\ell,j,i}}{\wh \sigma}\right)^2 + \sum_{i=1}^{p} \left(\frac{\gamma_{\ell,i}}{\wh \varsigma_i}\right)^2\right],
\end{equation}
where 
\begin{equation} \nonumber
\widetilde y_{\ell,j} = y_{\ell,j} - \sum_{i=1}^{p} x_{\ell,j,i} \wh \beta_i, \,\,\forall \, \ell = 1,\cdots,g; j = 1,\cdots,n_\ell.
\end{equation}
Notice also that the $\mathcal L_{\boldsymbol \gamma}(\boldsymbol \gamma)$ is only defined in the following set:
\begin{equation} \nonumber
\left\{\boldsymbol \gamma \, \left| \, -\wh{\boldsymbol \beta} \, \leq \, \boldsymbol \gamma^\ell \, \leq \,  \wh{\boldsymbol \beta}, \,\, \forall \, \ell = 1,\cdots,g  \right. \right\}.
\end{equation}
Therefore, maximizing (\oldref{eq:likelihood_gamma}) is equivalent to solving the following constrained optimization problem
\begin{equation}\label{eq:optimization_gamma}
\begin{array}{rl}
(\wh{\boldsymbol\gamma}^1,\cdots,\wh{\boldsymbol\gamma}^g) \, = \, \displaystyle{ \argmin_{\boldsymbol\gamma^1,\cdots,\boldsymbol\gamma^g} } & \displaystyle{ \sum_{\ell=1}^g \left[\frac{1}{\wh \sigma^2}\left(\widetilde{ \boldsymbol y}^\ell - Z^\ell \boldsymbol \gamma^\ell\right)^T\left(\widetilde{\boldsymbol y}^\ell - Z^\ell \boldsymbol \gamma^\ell\right) + \left(\boldsymbol \gamma^\ell\right)^T (\wh \Sigma)^{-1} \boldsymbol \gamma^\ell \right]} \\[10pt]
\mbox{s.t.}& -\wh{\boldsymbol \beta} \, \leq \, \boldsymbol \gamma^\ell \, \leq \,  \wh{\boldsymbol \beta}, \,\, \forall \, \ell = 1,\cdots,g,
\end{array}
\end{equation}
where
\begin{equation} \nonumber
\widetilde{\boldsymbol y}^\ell \, \triangleq \, \left[
\begin{array}{c}
\widetilde{y}_{\ell,1}\\
\vdots\\
\widetilde{y}_{\ell,n_\ell}
\end{array}\right] \, \in \, \mathbb R^{n_\ell}, \forall \, \ell=1,\cdots,g, \mbox{ and } \wh \Sigma \, \triangleq \,
\left[
\begin{array}{ccc}
\wh \varsigma_1^2\\
& \ddots \\
&& \wh \varsigma_k^2
\end{array}
\right].
\end{equation}
Note that (\oldref{eq:optimization_gamma}) is decomposable to solving for each $\boldsymbol \gamma^\ell$ independently. In fact, for each $\ell=1,\cdots,g$, we can solve
\begin{equation}\label{eq:optimization_gamma_individual}
\begin{array}{rl}
\wh{\boldsymbol\gamma}^\ell \, = \, \displaystyle{ \min_{\boldsymbol \gamma^\ell} } & \displaystyle{ \frac{1}{\wh \sigma^2}\left(\widetilde{ \boldsymbol y}^\ell - Z^\ell \boldsymbol \gamma^\ell\right)^T\left(\widetilde{\boldsymbol y}^\ell - Z^\ell \boldsymbol \gamma^\ell\right) + \left(\boldsymbol \gamma^\ell\right)^T (\wh \Sigma)^{-1} \boldsymbol \gamma^\ell } \\[10pt]
\mbox{s.t.}& -\wh{\boldsymbol \beta} \, \leq \, \boldsymbol \gamma^\ell \, \leq \,  \wh{\boldsymbol \beta}.
\end{array}
\end{equation}
For a linear mixed effects model, we follow \cite{nakagawa2013general} in defining the marginal $R^2$ and the conditional $ R^2 $ as follows.
\begin{equation}
\text{mar}_{R^2} = \left. \left(\sum_{j=1}^{n}(\hat{y}_j - \bar{y})^2/n\right)\right/\left(\left(\sum_{j=1}^{n}(\hat{y}_j - \bar{y})^2/n\right) + \sum_{i=1}^{k} \varsigma_k^2 + \sigma^2\right),
\end{equation}
and 
\begin{equation}
\text{con}_{R^2} = \left.\left(\sum_{j=1}^{n}(\hat{y}_j - \bar{y})^2/n  + \sum_{i=1}^{k} \varsigma_k^2 \right) \right/\left(\left(\sum_{j=1}^{n}(\hat{y}_j - \bar{y})^2/n \right) + \sum_{i=1}^{k} \varsigma_k^2 + \sigma^2\right).
\end{equation}
The marginal $ R^2 $ measures the proportion of the variance that the fixed effects can explain: the numerator is the variance of fixed effects, while the denominator is the total variance of the model, i.e., the sum of the variance of the fixed effects, variance of all random effects and variance of the error. In a similar fashion, conditional $R^2$ depicts the proportion of the variance that the whole regression model, i.e, both the fixed effects and random effects can explain. These two metrics are natural extensions of the usual $ R^2 $ to the mixed effects models, and they are between $0$ to $1$ inclusively.

\section{Simulation} \label{sec:simulation}
%We report some simulation results in this section.
\subsection{\textbf{Estimation}}
We compare the performance of the proposed methods with the unconstrained case (use the \textit{lme4} package in R) with different combinations of sample sizes. For each column of the design matrix $ X $, we independently simulate from a $\Gamma(2, 1) $ distribution with groups $ g=2 $. In the following tables, we name and abbreviate the proposed methods in Section \oldref{sec:methods} as  PLS (penalized least squares) and PRLS (penalized restricted least squared), respectively. We report the true parameters and estimates from \textit{lme4}, PLS and PRLS in Table \oldref{tab:intercept_p3} for LME with random intercept only for $p=3$, and in Table \oldref{tab:full_p3} with the LME with full random effects for $ p=3 $. Note that in this section, all models considered include an intercept term: for instance, when we talk about $ p=3 $, we mean $ \beta_0,\beta_1$ and $\beta_2 $ with $ \beta_0 $ as the intercept. In Table~\oldref{tab:intercept_p3}, the sample sizes considered are $ 300, 500, 1000 $. In Table~\oldref{tab:full_p3}, the sample sizes are $ 500, 1000, 2000 $. 

\begin{table}[htbp!]
	\begin{center}
		\caption {LME with random intercept only with $ p=3, g=2$.}  \label{tab:intercept_p3}
		%\hspace*{-1.5cm}
		\resizebox{\textwidth}{!}{\begin{tabular}{|c|c|ccc|ccc|ccc|}\hline
				&	& \multicolumn{3}{|c|}{$n=300$} & \multicolumn{3}{|c|}{$n=500$} & \multicolumn{3}{|c|}{$n=1000$} \\ \hline
				& True & \textit{lme4} & PLS & PRLS & \textit{lme4} & PLS & PRLS & \textit{lme4} & PLS & PRLS \\ \hline
				$ \beta_{1, 0} = \beta_0 + \gamma_{1, 0} $ & $ 0.000 $ & $ -0.171 $ & $0.000$ & $0.000$ &
				$ -0.061 $ & $ 0.000 $ & $0.000$ & $  0.030 $ &   $ 0.001 $ &  $ 0.001 $ \\
				
				$ \beta_{2, 0} = \beta_0 + \gamma_{2, 0} $ & $  0.144  $ & $ -0.005 $ & $  0.066 $ & $ 0.065 $ &
				$ -0.061 $ &  $ 0.013 $ & $ 0.006 $ & $ 0.106  $  & $ 0.136 $ & $ 0.136 $ \\
				
				$ \beta_1 $ & $ 1.000 $ & $ 1.069 $ & $ 1.044 $ & $ 1.044 $   &
				$ 1.002 $ & $ 0.987 $ & $0.990$ & $  0.993 $  & $ 0.992 $ &  $ 0.992 $ \\
				
				$  \beta_2 $  & $ 1.000 $ & $  1.013 $ & $  0.990 $ & $ 0.990 $ &
				$ 1.027 $ & $ 1.018 $ & $ 1.012 $ &  $ 1.021  $ & $ 1.021 $ & $ 1.021 $ \\
				
				$ s(\gamma_{i, 0}) $ & $ 0.058 $ & $ 0.136 $ & $  0.019 $ & $ 0.019 $  &
				$ 0.000 $ & $ 0.004 $ & $ 0.002 $ & $ 0.051  $  & $ 0.040 $ & $ 0.040 $ \\
				
				$  \sigma  $ & $ 1.000 $ & $ 0.976 $ & $ 0.983 $ & $  0.988 $  &
				$ 1.010 $ & $ 1.001 $ &  $ 1.014 $ & $ 1.000 $   & $ 1.001 $ & $ 1.002 $ \\ \hline
				
				RMSE &  & $ 0.102  $ & $ 0.041$ & $ 0.041 $  & $ 0.091 $ & $ 0.059 $ & $0.061$ &
				$ 0.022 $ & $ 0.012 $ & $ 0.012 $   \\ \hline
				
				Marginal $R^2$ &  & $ 0.825 $ & $ 0.818 $ & $ 0.817 $  & $ 0.802 $ & $ 0.870 $ & $ 0.867 $  &
				$ 0.793 $ & $ 0.927 $ &  $ 0.927 $ \\ \hline
				
				Conditional $R^2$ &  & $ 0.828 $ & $  0.818 $ & $ 0.817 $  & $ 0.802 $ & $ 0.870 $ & $ 0.867 $  &
				$ 0.793 $ & $ 0.927 $ &  $ 0.927 $ \\ \hline
		\end{tabular}}
	\end{center}
\end{table}	

\begin{table}[htbp!]
	\caption {LME with full random effects with $ p=3, g=2$.}  \label{tab:full_p3}
	\begin{center}
		%\hspace*{-1.5cm}
		\resizebox{\textwidth}{!}{\begin{tabular}{|c|c|ccc|ccc|ccc|}\hline
				&	& \multicolumn{3}{|c|}{$n=500$} & \multicolumn{3}{|c|}{$n=1000$} & \multicolumn{3}{|c|}{$n=2000$} \\ \hline
				& True & \textit{lme4} & PLS &PRLS & \textit{lme4} & PLS & PRLS & \textit{lme4} & PLS & PRLS \\ \hline
				$ \beta_{1, 0} = \beta_0 + \gamma_{1, 0} $ & $ 0.419 $ & $ 0.365 $ & $0.239$ & $0.238$ &
				$ 0.546 $ & $ 0.511 $ & $ 0.508 $
				& $ 0.370 $ &  $ 0.388 $ & $ 0.372 $ \\
				
				$ \beta_{2, 0} = \beta_0 + \gamma_{2, 0} $ & $  0.834  $ & $ 0.365 $ & $  0.585 $ &  $ 0.586 $  &
				$ 0.923 $ &  $ 0.968 $ & $ 0.970 $
				& $ 0.737 $ & $ 0.787 $ & $ 0.769 $ \\   \hline
				
				$ \beta_{1, 1} = \beta_1 + \gamma_{1, 1} $ & $ 0.644 $ & $ 0.665 $ & $0.679$ & $ 0.679  $  &
				$ 0.619 $ & $ 0.631 $ & $ 0.630 $
				& $ 0.662 $ &  $ 0.650 $ & $ 0.655 $ \\
				
				$ \beta_{2, 1} = \beta_1 + \gamma_{2, 1} $ & $  0.374  $ & $ 0.420 $ & $  0.367 $ & $ 0.367 $ &
				$ 0.317 $ &  $ 0.320 $ &  $ 0.316 $
				& $ 0.403 $ & $ 0.376 $ & $ 0.391 $ \\  \hline
				
				$ \beta_{1, 2} = \beta_2 + \gamma_{1, 2} $ & $ 1.108 $ & $ 1.119 $ & $1.157$ & $ 1.157 $ &
				$ 1.076 $ & $ 1.079 $ & $ 1.082 $
				& $ 1.105 $ &  $ 1.105 $ & $ 1.108 $ \\
				
				$ \beta_{2, 2} = \beta_2 + \gamma_{2, 2} $ & $  0.926  $ & $ 1.040 $ & $  0.996 $ & $ 0.995 $  &
				$ 0.930 $ &  $ 0.909 $ & $ 0.911 $
				& $ 0.937 $ & $ 0.935 $ & $ 0.932 $ \\  \hline
				
				$ s(\gamma_{i, 0}) $ & $ 0.540 $ & $ 0.003 $ & $  0.225 $ &  $ 0.233 $ &
				$ 0.283 $ & $ 0.411 $ & $ 0.352 $
				& $ 0.268 $ & $ 0.178 $ & $ 0.321 $ \\
				
				$ s(\gamma_{i, 1}) $ & $ 0.540 $ & $ 0.176 $ & $  0.170 $ & $ 0.173 $ &
				$ 0.216 $ & $ 0.183 $ &  $ 0.144 $
				& $ 0.184 $ & $ 0.249 $ & $ 0.144 $ \\
				
				$ s(\gamma_{i, 2}) $ & $ 0.540 $ & $ 0.064 $ & $  0.051 $ & $ 0.060 $ &
				$ 0.108 $ & $ 0.038 $ & $ 0.059 $
				& $ 0.121 $ & $ 0.180 $ & $ 0.112 $ \\ \hline
				
				$  \sigma  $ & $ 1.000 $ & $ 0.971 $ & $ 0.909 $ & $ 0.908 $ &
				$ 1.009 $ & $ 0.852 $ & $ 0.936 $
				& $ 0.968 $ & $ 0.856 $ & $ 0.868 $ \\ \hline
				
				RMSE &  & $ 0.299  $ & $ 0.243$ & $ 0.238 $ &
				$ 0.196 $ & $ 0.211 $ & $ 0.214 $ &
				$ 0.198 $ & $ 0.192 $ &  $ 0.203 $ \\ \hline
				
				Marginal $R^2$ &  & $ 0.761 $ & $  0.769 $ &  $ 0.769 $ &
				$ 0.666 $ & $ 0.697 $ & $ 0.702 $ &
				$ 0.710 $ & $ 0.771  $ &  $ 0.771 $ \\ \hline
				
				Conditional $R^2$ &  & $ 0.770 $ & $  0.791 $ &  $ 0.791 $ &
				$ 0.706 $ & $ 0.780 $ & $ 0.777 $ &
				$ 0.743 $ & $ 0.833  $ &  $ 0.835 $ \\ \hline
		\end{tabular}}
	\end{center}
\end{table}

\begin{figure}[htbp!]
	\hspace*{-0.5cm}
	\subfloat{{\includegraphics[width=0.5\textwidth, height=5.85cm]{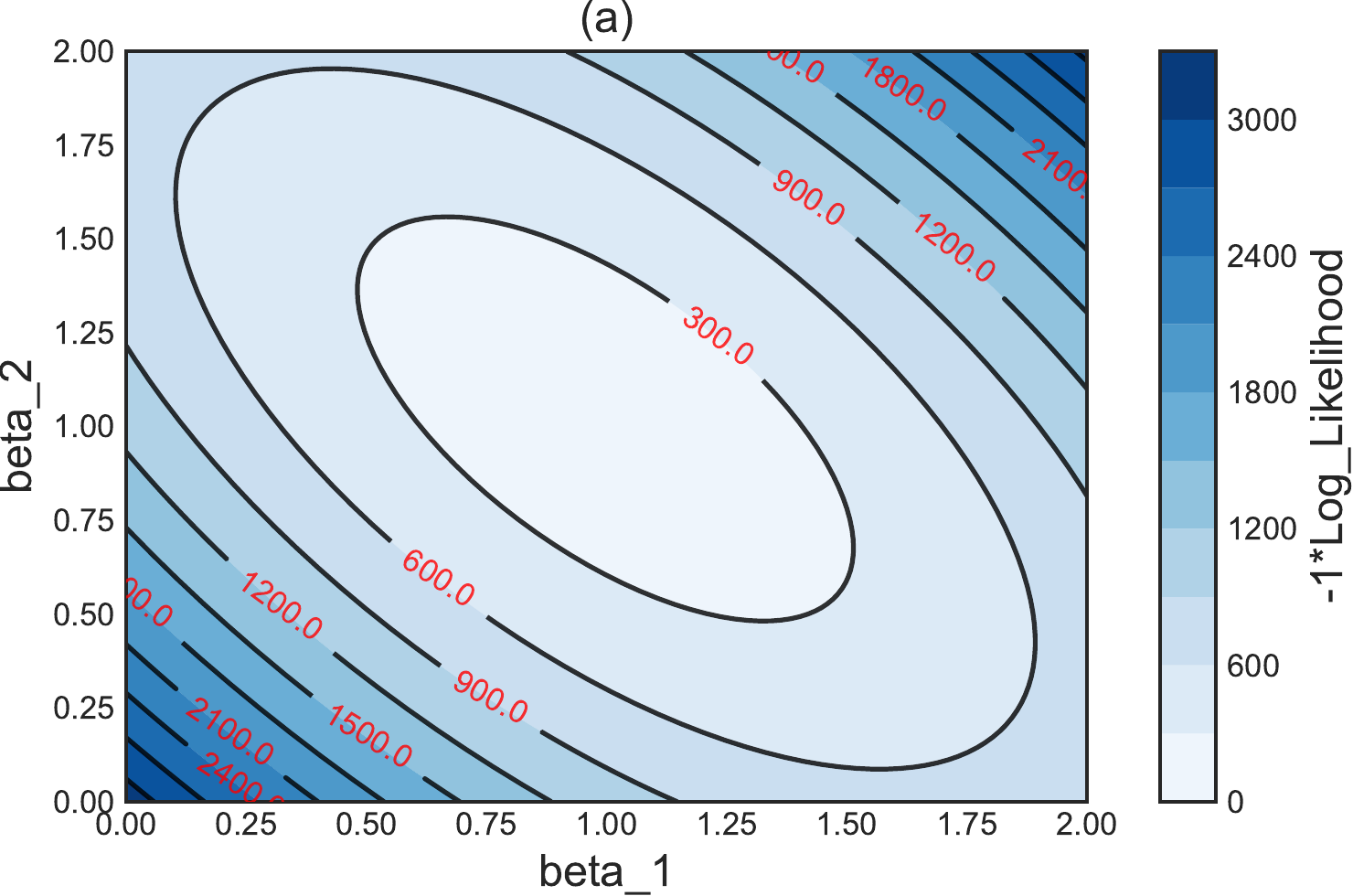}}}%
	\qquad
	\subfloat{{\includegraphics[width=0.5\textwidth, height=5.85cm]{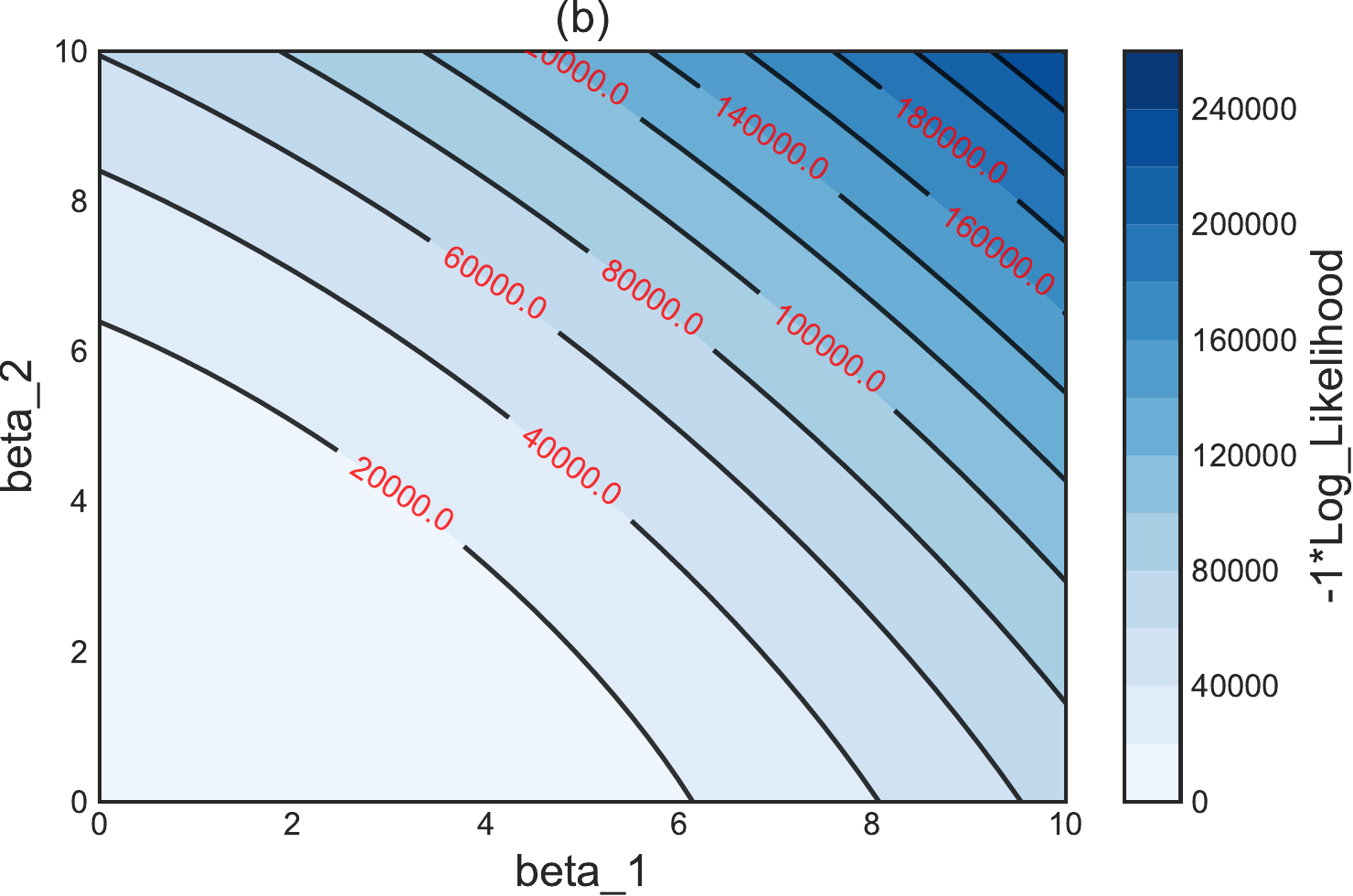}}}%
	\caption{Contour plots of the log-likelihood function in equation (\oldref{eqn:pls}) viewed as a function of $\beta_1$ and $ \beta_2 $. The data range for (a) is $ 0 $ to $ 2 $, while the range for (b) is $0$ to $10$.}%
	\label{fig:con1}%
\end{figure}

\begin{figure}[htbp!]
	\hspace*{-0.5cm}
	\subfloat{{\includegraphics[width=0.5\textwidth, height=5.85cm]{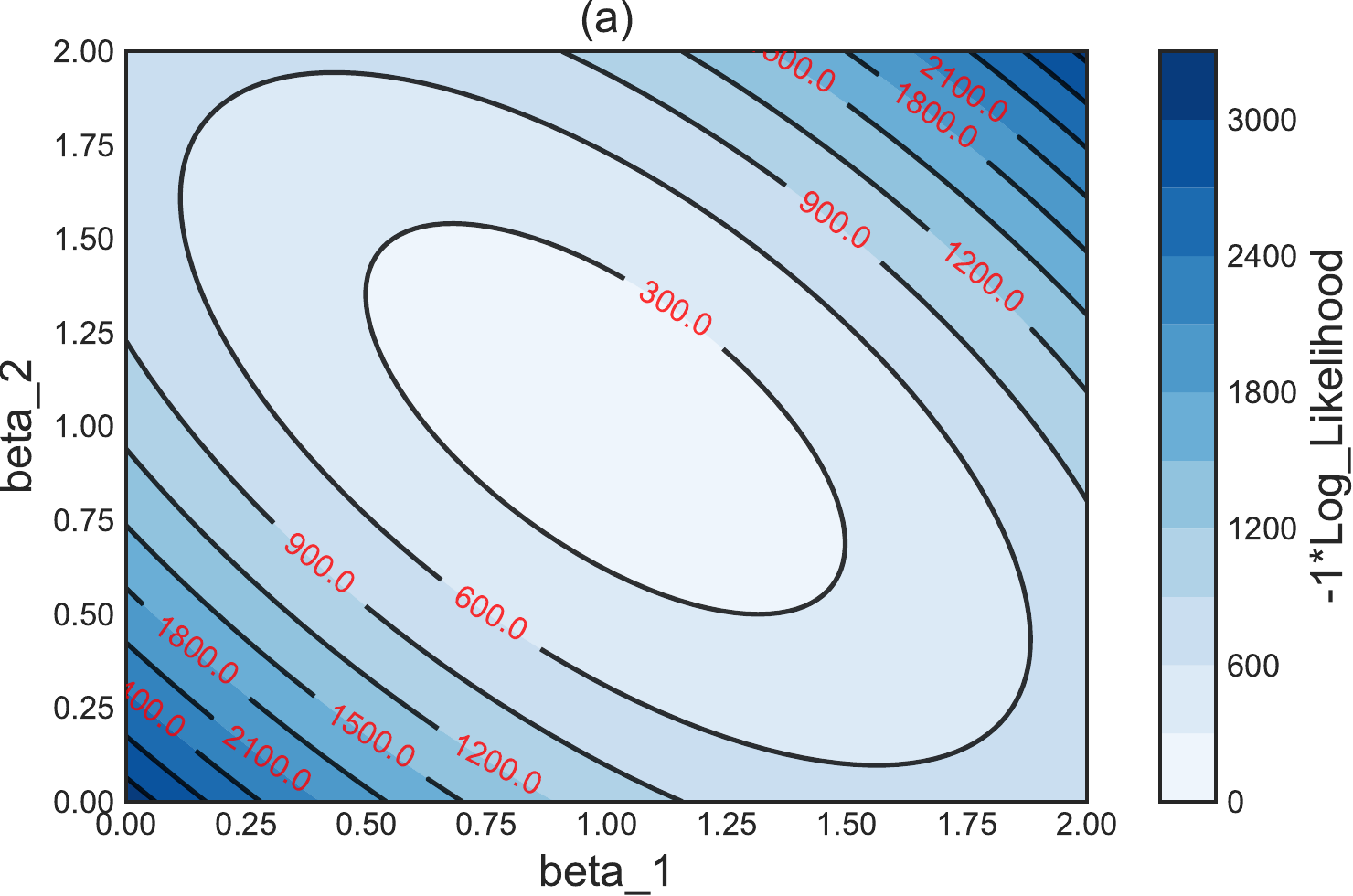}}}%
	\qquad
	\subfloat{{\includegraphics[width=0.5\textwidth, height=5.85cm]{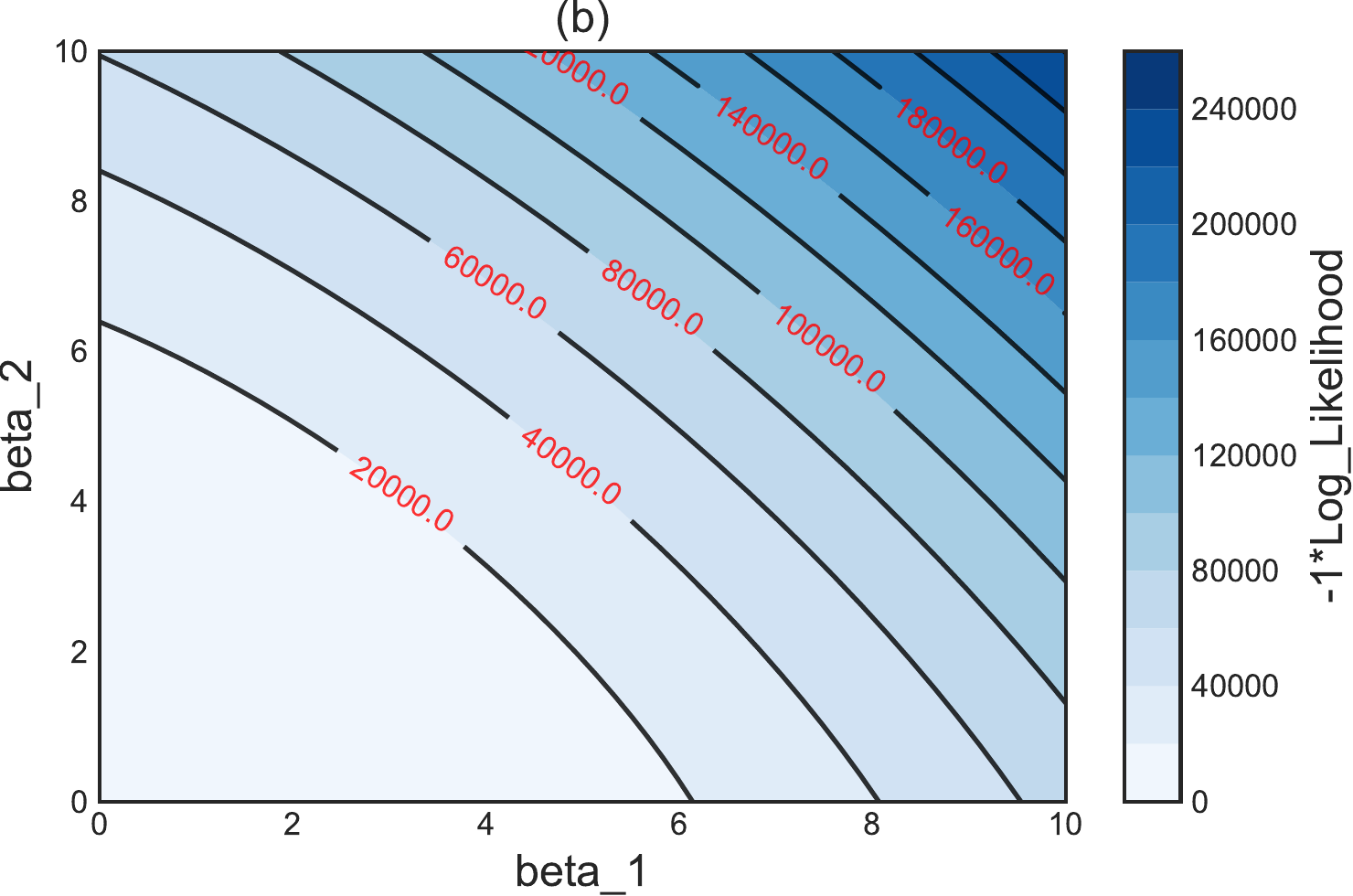}}}%
	\caption{Contour plots of the log-likelihood function in equation (\oldref{eqn:reML}) viewed as a function of $\beta_1$ and $ \beta_2 $. The data range for (a) is $ 0 $ to $ 2 $, while the range for (b) is $0$ to $10$.}%
	\label{fig:con2}%
\end{figure}

Before analyzing the simulation results, we first examine the contour plots in Figures~\oldref{fig:con1} and \oldref{fig:con2}, exhibiting the behavior of the objective function in equation (\oldref{eqn:pls}) and in equation (\oldref{eqn:reML}), respectively, which is viewed as a function of $\beta_1$ and $ \beta_2 $. All other parameters are fixed at their true values. It is clear that both equations are unimodal such that the optimal value is obtained when $ \beta_1 = \beta_2 =1 $, which are the true parameters used to simulate the data. In addition, we also present figures that show behavior of the objective function in equation (\oldref{eqn:pls}) and in equation (\oldref{eqn:reML}) as a function of $ \beta_0 $ and $ \beta_1 $ in Section~C of the \textit{Supplemental Document}. They basically tell the same story as Figures \oldref{fig:con1} and \oldref{fig:con2}. Although the objective function is unimodal given the data ranges specified in the figures, we still run with $ 5 $ different initial values and the results with the lowest value are retained and reported in this section.  

As we analyze the simulation results reported in Tables \oldref{tab:intercept_p3} and \oldref{tab:full_p3}, note that $ s(\gamma_{i, 0}) $ in Table~\oldref{tab:intercept_p3} is standard derivation of the random effects. For a SDTN, it is $  \sqrt{\eta^2(1-\frac{2\rho\phi(\rho)}{2\Phi(\rho)-1})} $, where $ \rho =\beta/\eta$. From Table~\oldref{tab:intercept_p3}, we observe that the RMSE of PLS and PRLS are close to each other, and both are smaller than that of \textit{lme4}. In Table~\oldref{tab:intercept_p3}, it is obvious that both are about half of that for \textit{lme4}, for example, with the random intercept only model for $ n=500, p=3$, the RMSE of PLS and PRLS are $0.059$ and $0.061$, respectively, which is much lower than the $0.091$ of \textit{lme4}. From Table~\oldref{tab:full_p3}, the RMSE of PLS and PRLS are lower than \textit{lme4} for $n=500$, but are a bit higher for $ n=1000 $. The difference in RMSE is rather small.

Another important observation is that in Table~\oldref{tab:intercept_p3}, when $n=300$, the estimated intercepts of \textit{lme4} are negative when the true parameters are $0.000$ for group $1$ and $0.144$ for group 2. For PLS and PRLS, both estimates are non-negative. This is a convincing example to show the advantage of constraining the random effects via a SDTN. The results from the proposed methods are ensured to be consistent to the domain knowledge, making it more practical to use in reality than the unconstrained case. Moreover, these results were obtained with little to no sacrifice on the RMSE.  In other words, PLS and PRLS approaches were able to identify alternative parameter estimates that match our prior knowledge better with little compromise on model fit or obtaining an even better fit. 

Moreover, the results on marginal $R^2$ and conditional $ R^2 $ are interesting: the conditional $ R^2 $ are good for all the three methods indicating that the linear mixed effects model explains most of the total variance. In Table~\oldref{tab:intercept_p3}, the marginal $ R^2 $ is very close to conditional $ R^2 $, and it makes sense as the true variance of the random effect is rather small, $ s^2(\gamma_i, 0) = 0.058^2 $ compared to the variance of the error term, $ \sigma^2 = 1^2 $. In a clear contrast, as we consider random effects on more variables, the true $ s^2  $ becomes $  0.540^2 \times 3 $ in Table \oldref{tab:full_p3}, and the difference between marginal $ R^2 $ and conditional $ R^2 $ become more obvious. In Table \oldref{tab:intercept_p3}, PLS and PRLS have obvious higher marginal $R^2$ and conditional $R^2$ for $n=500, 1000$. In a similar pattern, Table \oldref{tab:full_p3} shows that PLS and PRLS perform better than \textit{lme4} on both marginal $R^2$ and conditional $R^2$. Overall, the proposed methods are able to explain more of the variance. Now, with the same setting, we increase the dimensions from $ 3 $ to $ 7 $ and report the results in Table~\oldref{tab:intercept_p7} for a model with random intercept only and Table~\oldref{tab:full_p7} for a model with full random effects.
\begin{table}[htbp!]
	\caption {LME with random intercept only with $ p=7, g=2$.}  \label{tab:intercept_p7}
	\begin{center}
		%\hspace*{-1.5cm}
		\resizebox{\textwidth}{!}{\begin{tabular}{|c|c|ccc|ccc|ccc|}\hline
				&	& \multicolumn{3}{|c|}{$n=1000$} & \multicolumn{3}{|c|}{$n=2000$} & \multicolumn{3}{|c|}{$n=4000$} \\ \hline
				& True & \textit{lme4} & PLS & PRLS  & \textit{lme4} & PLS & PRLS & \textit{lme4} & PLS & PRLS \\ \hline
				$ \beta_{1, 0} = \beta_0 + \gamma_{1, 0} $ & $ 0.157 $ & $ 0.157 $ & $0.219$ & $ 0.219 $ &
				$ 0.092 $ & $ 0.124 $ & $ 0.124 $  & $ 0.085 $ &  $ 0.125 $ & $ 0.125 $ \\
				
				$ \beta_{2, 0} = \beta_0 + \gamma_{2, 0} $ & $  1.948  $ & $ 2.019 $ & $  2.075 $ & $ 2.075 $ &
				$ 1.940 $ &  $ 1.969 $ & $1.969 $  &  $ 1.927 $ & $ 1.965 $ & $ 1.965 $ \\
				
				$  \beta_1 $  & $ 1.000 $ & $  0.975 $ & $  0.960 $ & $ 0.960 $ &
				$ 1.018 $ & $ 1.026 $ & $1.026 $  &  $ 0.976 $ & $ 0.984 $ & $ 0.984 $ \\
				
				$ \beta_2 $ & $ 1.000 $ & $ 1.048 $ & $ 1.068 $ & $ 1.068 $ &
				$ 1.008 $ & $ 0.996 $ & $ 0.996 $  & $ 1.029 $ &  $ 1.042 $ & $ 1.042 $ \\
				
				$  \beta_3 $  & $ 1.000 $ & $  1.026 $ & $  0.999 $ & $ 0.999 $ &
				$ 1.008 $ & $ 1.025 $ & $ 1.025 $ & $ 1.008 $ & $ 0.987 $ & $ 0.988 $ \\
				
				$ \beta_4 $ & $ 1.000 $ & $ 1.006 $ & $ 1.025 $ & $ 1.025 $ &
				$ 0.987 $ & $ 0.963 $ & $ 0.963 $ & $ 1.009 $ &  $ 1.000 $ &  $ 1.001 $ \\
				
				$  \beta_5 $  & $ 1.000 $ & $  0.966 $ & $  0.955 $ & $  0.955 $ &
				$ 0.987 $ & $ 0.978 $ & $ 0.978 $ & $ 1.025 $ & $ 1.018 $ & $ 1.018 $ \\
				
				$ \beta_6 $ & $ 1.000 $ & $ 0.958 $ & $ 0.944 $ & $ 0.944 $ &
				$ 1.009 $ & $ 1.013 $ & $ 1.013 $   & $ 0.986 $ &  $ 0.979 $ & $ 0.979 $ \\
				
				$ s(\gamma_{i, 0}) $ & $ 0.540 $
				& $ 0.932 $ & $  0.640 $ & $ 0.632 $
				& $ 0.924 $ & $ 0.575 $ & $ 0.575 $
				& $ 0.921 $ & $ 0.450 $ & $ 0.451 $ \\
				
				$  \sigma  $ & $ 1.000 $
				& $ 0.974 $ & $ 1.185 $ & $ 1.194 $ &
				$ 1.020 $ & $ 1.249 $ &  $ 1.252 $ &
				$ 0.998 $ & $ 1.281 $ &  $ 1.282 $ \\ \hline
				
				RMSE &  & $ 0.129  $ & $ 0.087$ & $ 0.088 $ &
				$ 0.124 $ & $ 0.083 $ & $ 0.083 $ &
				$ 0.124 $ & $ 0.095 $ &  $ 0.096 $ \\ \hline
				
				Marginal $R^2$ &  & $ 0.870 $ & $  0.870 $ &  $ 0.870 $ &
				$ 0.867 $ & $ 0.867 $ & $ 0.866 $ &
				$ 0.867 $ & $ 0.866  $ &  $ 0.866 $ \\ \hline
				
				Conditional $R^2$ &  & $ 0.902 $ & $  0.891 $ &  $ 0.896 $ &
				$ 0.907 $ & $ 0.891 $ & $ 0.891 $ &
				$ 0.908 $ & $ 0.881  $ &  $ 0.880 $ \\ \hline
		\end{tabular}}
	\end{center}
\end{table}	

%Let us do a full model with $ p=7 $.
\begin{table}[htbp!]
	\caption {LME with full random effects with $ p=7, g=2$.}  \label{tab:full_p7}
	\begin{center}
		%\hspace*{-1.5cm}
		\resizebox{\textwidth}{!}{\begin{tabular}{|c|c|ccc|ccc|ccc|}\hline
				&	& \multicolumn{3}{|c|}{$N=1000$} & \multicolumn{3}{|c|}{$N=2000$} & \multicolumn{3}{|c|}{$N=4000$} \\ \hline
				& True & \textit{lme4} & PLS & PRLS & \textit{lme4} & PLS & PRLS & \textit{lme4} & PLS & PRLS \\ \hline
				$ \beta_{1, 0} = \beta_0 + \gamma_{1, 0} $ & $ 1.520 $ & $ 1.386 $ & $1.425$ & $1.427$ &
				$ 1.423 $ & $ 1.290 $ & $1.291$ & $ 1.375 $ &  $ 1.361 $ & $1.361$  \\
				
				$ \beta_{2, 0} = \beta_0 + \gamma_{2, 0} $ & $  1.372  $ & $ 1.386 $ & $  1.412 $ & $  1.359 $ &
				$ 1.423 $ &  $ 1.522 $ & $  1.522 $ & $ 1.375 $ & $ 1.319 $ & $  1.345 $  \\
				
				$ \beta_{1, 1} = \beta_1 + \gamma_{1, 1} $ & $ 1.052 $ & $ 1.010 $ & $1.008$ & $  1.008 $ &
				$ 1.080 $ & $ 1.078 $ & $  1.079 $ & $ 1.045 $ &  $ 1.055 $ & $  1.041 $  \\
				
				$ \beta_{2, 1} = \beta_1 + \gamma_{2, 1} $ & $  0.739  $ & $ 0.803 $ & $ 0.815 $ & $  0.829 $ &
				$ 0.743 $ &  $ 0.724 $ & $  0.724 $ & $ 0.762 $ & $ 0.764 $ & $  0.762 $ \\
				
				$ \beta_{1, 2} = \beta_2 + \gamma_{1, 2} $ & $ 1.324 $ & $ 1.363 $ & $1.351$ & $  1.350 $ &
				$ 1.323 $ & $ 1.340 $ & $  1.342 $ & $ 1.357 $ &  $ 1.363 $ & $  1.355 $  \\
				
				$ \beta_{2, 2} = \beta_2 + \gamma_{2, 2} $ & $  1.680  $ & $ 1.697 $ & $  1.696 $ & $  1.697 $ &
				$ 1.703 $ &  $ 1.686 $ & $  1.698 $ & $ 1.673 $ & $ 1.698 $ & $  1.702 $  \\
				
				$ \beta_{1, 3} = \beta_2 + \gamma_{1, 2} $ & $ 1.194 $ & $ 1.202 $ & $1.184$ & $  1.185 $ &
				$ 1.183 $ & $ 1.174 $ & $  1.182 $ & $ 1.184 $ &  $ 1.177 $ & $ 1.190 $  \\
				
				$ \beta_{2, 3} = \beta_2 + \gamma_{2, 2} $ & $  1.342  $ & $ 1.315 $ & $  1.325 $ & $  1.333 $ &
				$ 1.367 $ &  $ 1.359 $ & $  1.362 $ & $ 1.339 $ & $ 1.343 $ & $  1.339 $  \\
				
				$ \beta_{1, 4} = \beta_2 + \gamma_{1, 2} $ & $ 1.153 $ & $ 1.177 $ & $1.171$ & $  1.169 $ &
				$ 1.132 $ & $ 1.165 $ & $  1.164 $ & $ 1.130 $ &  $ 1.136 $ & $  1.142 $  \\
				
				$ \beta_{2, 4} = \beta_2 + \gamma_{2, 2} $ & $  0.291  $ & $ 0.300 $ & $  0.268 $ & $  0.269 $ &
				$ 0.282 $ &  $ 0.271 $ & $  0.282 $ & $ 0.280 $ & $ 0.263 $ & $  0.260 $  \\
				
				$ \beta_{1, 5} = \beta_2 + \gamma_{1, 2} $ & $ 0.534 $ & $ 0.548 $ & $0.539$ & $  0.539 $ &
				$ 0.563 $ & $ 0.574 $ & $  0.576 $ & $ 0.577 $ &  $ 0.572 $ & $  0.562 $  \\
				
				$ \beta_{2, 5} = \beta_2 + \gamma_{2, 2} $ & $  0.609  $ & $ 0.614 $ & $  0.612 $ & $  0.607 $ &
				$ 0.566 $ &  $ 0.570 $ & $  0.565 $ & $ 0.594 $ & $ 0.601 $ & $  0.599 $  \\
				
				$ \beta_{1, 6} = \beta_2 + \gamma_{1, 2} $ & $ 0.559 $ & $ 0.556 $ & $0.579$ & $  0.580 $ &
				$ 0.571 $ & $ 0.588 $ & $  0.577 $ & $ 0.597 $ &  $ 0.592 $ & $  0.604 $  \\
				
				$ \beta_{2, 6} = \beta_2 + \gamma_{2, 2} $ & $  1.101  $ & $ 1.071 $ & $  1.072 $ & $  1.082 $ &
				$ 1.100 $ &  $ 1.101 $ & $  1.099 $ & $ 1.104 $ & $ 1.112 $ & $  1.103 $  \\  \hline

				$ s(\gamma_{i, 0}) $ & $ 0.540 $ & $ 0.001 $ & $  0.583 $ & $  0.436 $ &
				$ 0.000 $ & $ 0.108 $ & $  0.109 $ & $ 0.000 $ & $ 0.474 $ & $  0.508 $ \\
				
				$ s(\gamma_{i, 1}) $ & $ 0.540 $ & $ 0.149 $ & $  0.381 $ & $  0.369 $ &
				$ 0.239 $ & $ 0.018 $ & $  0.026 $ & $ 0.200 $ & $ 0.595 $ & $  0.389 $  \\
				
				$ s(\gamma_{i, 2}) $ & $ 0.540 $ & $ 0.238 $ & $  0.215 $ & $  0.191 $ &
				$ 0.269 $ & $ 0.087 $ & $  0.089 $ & $ 0.226 $ & $ 0.497 $ & $  0.362 $ \\
				
				$ s(\gamma_{i, 3}) $ & $ 0.540 $ & $ 0.084 $ & $  0.077 $ & $  0.130 $ &
				$ 0.132 $ & $ 0.606 $ & $  0.606 $ & $ 0.110 $ & $ 0.527 $ & $  0.114 $  \\
				
				$ s(\gamma_{i, 4}) $ & $ 0.540 $ & $ 0.620 $ & $  0.524 $ & $  0.384 $ &
				$ 0.601 $ & $ 0.437 $ & $  0.436 $ & $ 0.594 $ & $ 0.967 $ & $  0.301 $  \\
				
				$ s(\gamma_{i, 5}) $ & $ 0.540 $ & $ 0.053 $ & $  0.407 $ & $  0.248 $ &
				$ 0.006 $ & $ 0.312 $ & $  0.312 $ & $ 0.016 $ & $ 0.288 $ & $  0.276 $  \\
				
				$ s(\gamma_{i, 6}) $ & $ 0.540 $ & $ 0.365 $ & $  0.386 $ & $  0.347 $ &
				$ 0.374 $ & $ 0.396 $ & $  0.397 $ & $ 0.360 $ & $ 1.025 $ & $  0.205 $  \\ \hline
				
				$  \sigma  $ & $ 1.000 $ & $ 0.981 $ & $ 1.317 $ & $ 0.883   $ &
				$ 0.999 $ & $ 0.838 $ & $  0.844 $ & $ 1.010 $ & $ 1.296 $ & $  1.036 $  \\ \hline
				
				RMSE &  & $ 0.218  $ & $ 0.152$ & $ 0.153 $ &
				$ 0.209 $ & $ 0.198 $ & $ 0.196 $ &
				$ 0.216 $ & $ 0.167 $ &  $ 0.152 $ \\ \hline
				
				Marginal $R^2$ &  & $ 0.896 $ & $  0.810 $ &  $ 0.807 $ &
				$ 0.887 $ & $ 0.905 $ & $ 0.904 $ &
				$ 0.889 $ & $ 0.906  $ &  $ 0.906 $ \\ \hline
				
				Conditional $R^2$ &  & $ 0.926 $ & $  0.908 $ &  $ 0.906 $ &
				$ 0.927 $ & $ 0.958 $ & $ 0.957 $ &
				$ 0.930 $ & $ 0.951  $ &  $ 0.950 $ \\ \hline
		\end{tabular}}
	\end{center}
\end{table}
From Tables~\oldref{tab:intercept_p7} and \oldref{tab:full_p7}, we make similar observations to those of the previous two tables with $p=3$: compared to the unconstrained method, the proposed methods not only preserve interpretability, but also have satisfactory performance: the RMSEs are lower than \textit{lme4} for all the combinations considered. The model fits are also comparable with \textit{lme4} even if the true parameters are not so close to $ 0 $. These observations are consistent with previous results, demonstrating the usefulness of the new methods.

\subsection{\textbf{Merits of Sign Constraints}}
\begin{figure}[htbp!]
	%\centering
	\hspace*{-0.7cm}
	\subfloat{{\includegraphics[width=0.5\linewidth, height=5.85cm]{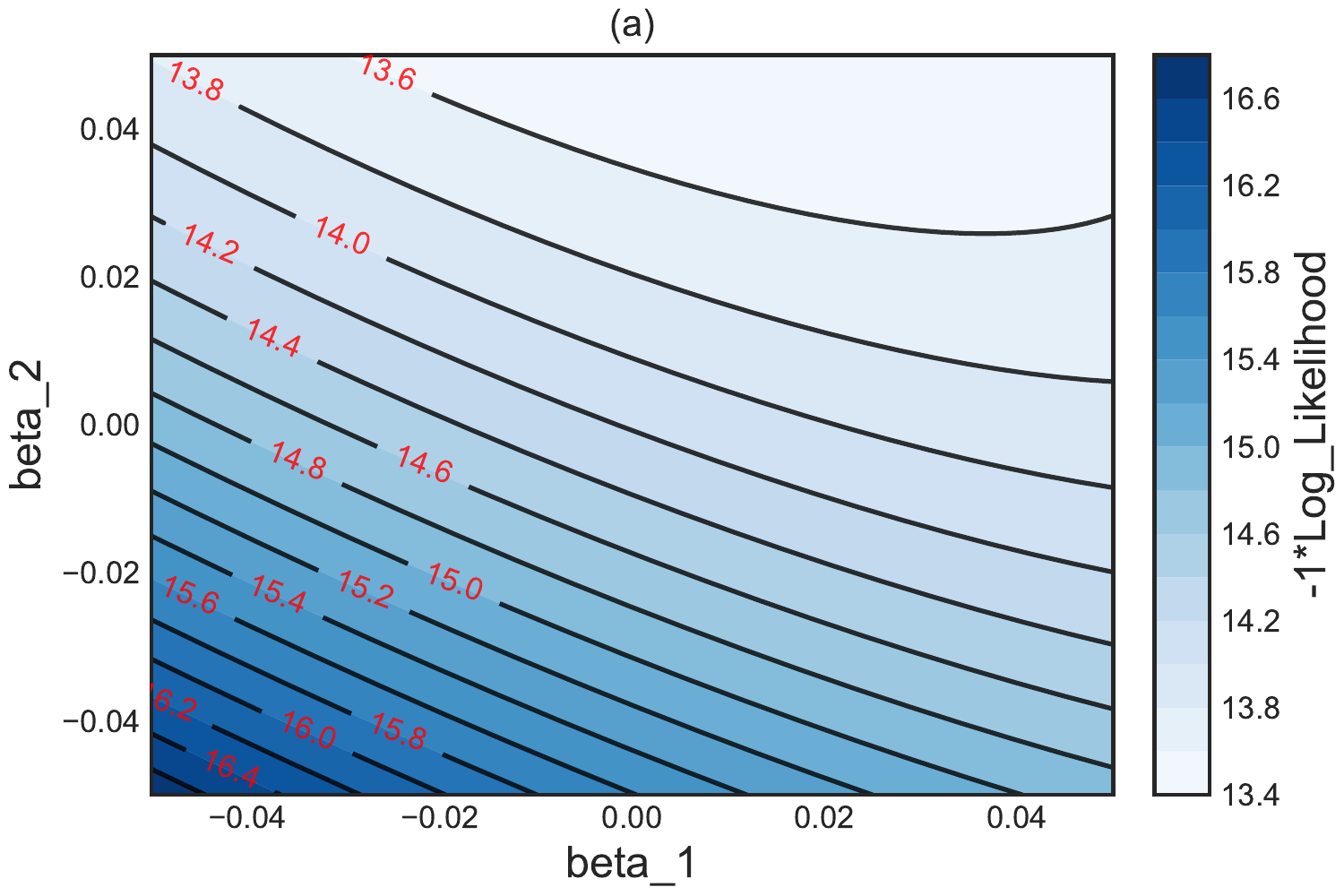}}}%
	\qquad
	\subfloat{{\includegraphics[width=0.5\linewidth, height=5.85cm]{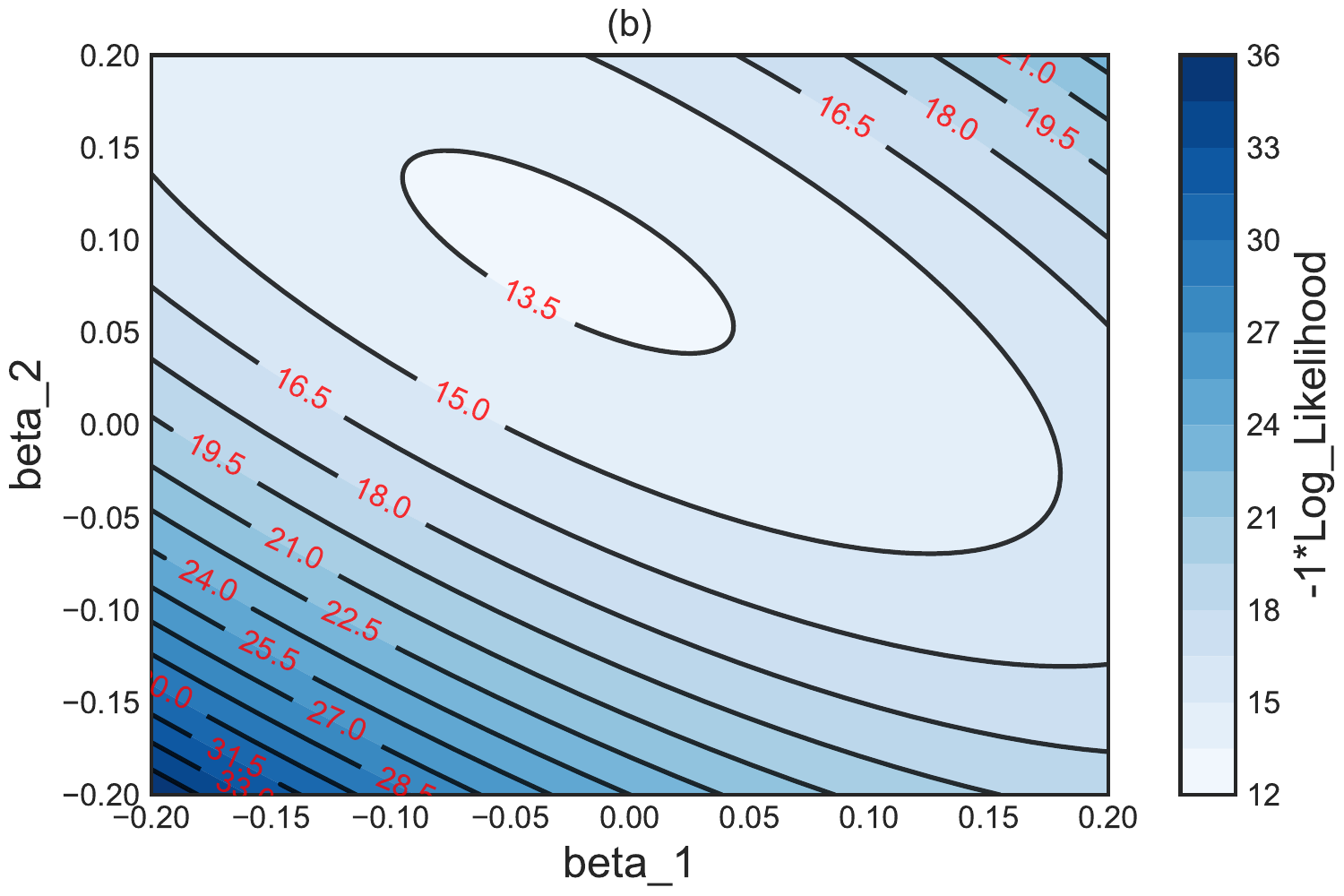}}}%
	\caption{Contour plots of the log-likelihood function in equation (\oldref{eqn:pls}) viewed as a function of $\beta_1$ and $ \beta_2 $. The data range for (a) is $ -0.05 $ to $ 0.05 $, while the range for (b) is $-0.02$ to $0.02$.}%
	\label{fig:con15}
\end{figure}

\begin{figure}[htbp!]
	\centering
	%\hspace*{-1.0cm}
	\includegraphics[width=8.0cm, height=5.85cm]{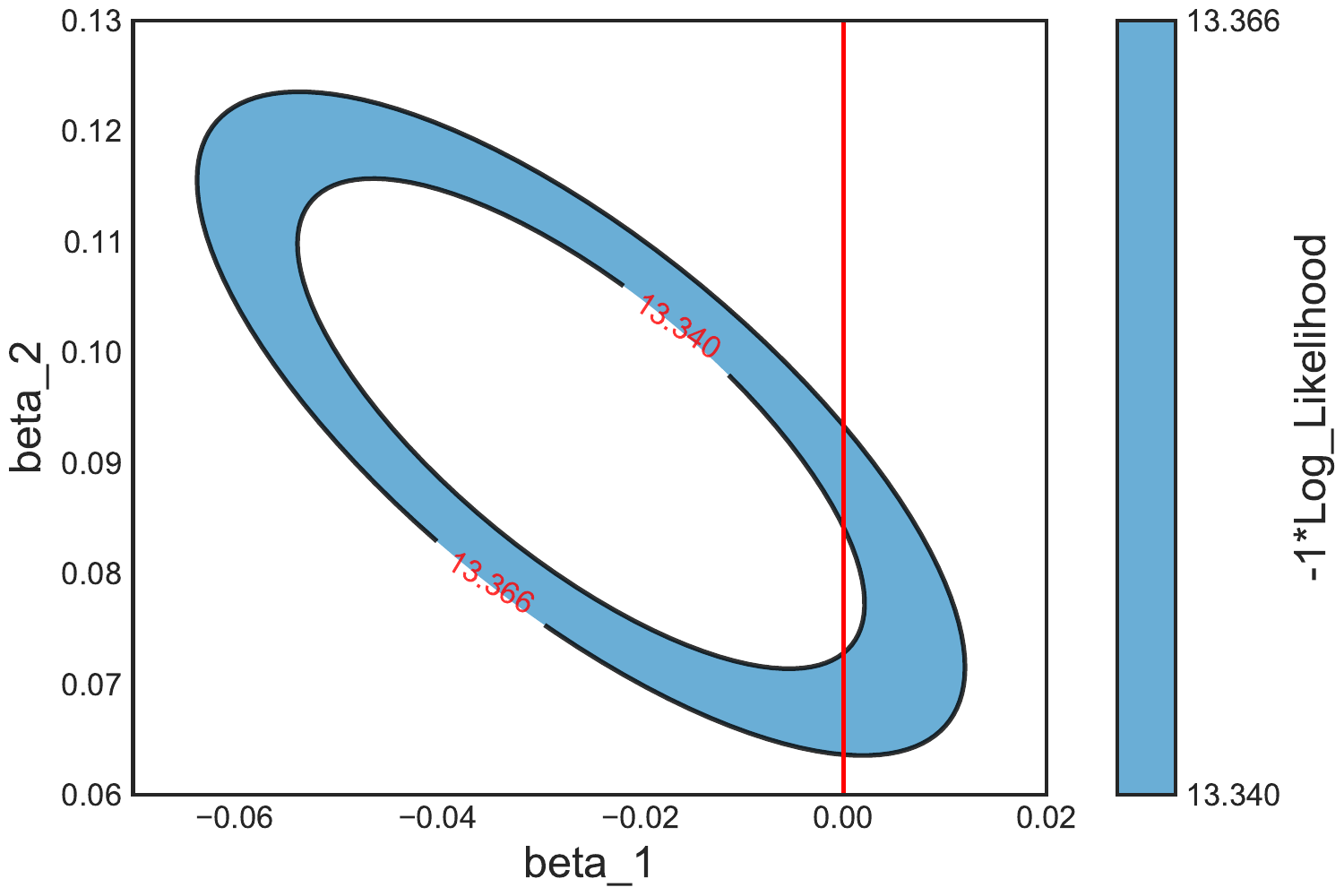}%
	%\subfloat{{\includegraphics[width=8.0cm, height=5.85cm]{more_plot_020.pdf}}}%
	\caption{Contour lines of the log-likelihood function in equation (\oldref{eqn:pls}) viewed as a function of $\beta_1$ and $ \beta_2 $. Two contour lines are specifically plotted: $13.340$ and $13.366$.}%
	\label{fig:con16}
\end{figure}

As has been stressed in previous sections, the sign constraints on the regression parameters not only lead to practical benefits such that resulting estimates comply with business knowledge automatically, but also bear theoretical merits that yield more accurate estimates by shrinking feasible parameter space. We still stick to $ p=3, g=2 $, and the objective function in equation (\oldref{eqn:pls}) is deemed as a function of $ \beta_1, \beta_2$ to facilitate producing contour plots. The true values for $ \beta_1 $ and $ \beta_2 $ are chosen as $ 0.001 $, which are close to $ 0 $. All other parameters are fixed at their true values. Sample size $ n $ is chosen as $ 30 $. Some estimation results are reported in Table~\oldref{tab:con15}. The contour plots are presented in Figures~\oldref{fig:con15} and \oldref{fig:con16}. 

It is clear from Table~\oldref{tab:con15} that both estimations find better results as measured by the objective value of equation (\oldref{eqn:pls}), for which the lower, the better. Due to small sample size, $n=30$ used to simulate data, variation is relatively large, and it is not unexpected that although both $ \beta_1 = 0.001 $ and $ \beta_2 = 0.001 $ are positive, the estimates without any constraint end up with $\hat{\beta_1} = -0.026$ that yields the lowest objective value, $13.340$. However, with non-negative sign constraints on both parameters, it actually shrinks the feasible parameter space, leading to a similar objective value, $ 13.366 $. More importantly, it ``corrects" the unconstrained estimate of $ \beta_1 $ to $ 0.000 $, while leaving $ \beta_2 $ largely unaffected. The contour plots in Figure \oldref{fig:con15} confirm the observation. 

In addition, we present Figure \oldref{fig:con16}, in which two contour lines, $13.340$ and $13.366$ are specifically drawn. We also add a vertical line $ \beta_1 = 0 $ in red to represent the boundary for $ \beta_1$ in the same figure. For contour line $ 13.340 $, it is apparent that there does exist regions of $(\beta_1, \beta_2)$ that comply with non-negative sign constraints, but the unconstrained algorithm was not able to arrive at any of those solutions.  Instead it reported an alternative solution with wrong sign for $\beta_1$ of $ (-0.026, 0.109) $ as reported in Table \oldref{tab:con15}. Now, with the help of sign constraints, the constrained optimization picked $(0.000, 0.093)$, which is an even more convincing merit of sign constraints. In other words, with a small sample size and close-to-boundary true parameters, the merits of imposing sign constraints on the regression parameters allowed the selection of alternative solutions that conforms with interpretability of the parameters.   

\begin{table}[htbp!]
\centering
\caption{Estimation results of unconstrained case and constrained case. The lower the objective value is, the better.}
\resizebox{\textwidth}{!}{
\begin{tabular}{|l|c|c|c|} \hline
& $\beta_1$  & $\beta_2$ & Objective value of Equation (\oldref{eqn:pls}) \\ \hline
True parameters plugged-in  & $0.001$  & $0.001$ & $14.152$    \\ \hline
Estimation without sign constraints & $-0.026$ & $0.109$ & $13.340$  \\ \hline
Estimation with non-negative sign constraints & $0.000$  & $0.093$ & $13.366$ \\ \hline 
\end{tabular}
}
\label{tab:con15}
\end{table}

\subsection{\textbf{Comparisons with the Method of PIT}}  \label{sec:pit}
It is interesting to directly compare the performance of the proposed methods to the method of PIT \citep{nelson2006use}, which has been briefly reviewed in Section~\oldref{sec:intro}. Both the proposed methods and the PIT method are able to handle LME models with non-Normal random effects. First of all, the proposed approaches are discriminative in nature, approximating the marginal distributions of the response directly in estimating fixed effects, while the PIT approach is a generative method that depends on the joint distribution of random effects and response variable. Secondly, both methods work best with independent covariance structure on random effects, and both require some non-trivial future work in order to deal with random effects with dependent covariance structure. Thirdly, it is easier using the proposed methods to incorporate more than one random effect as demonstrated in previous sections, while the PIT method will require the approximation of multiple integrals, in which the so-called ``curse of dimensionality" could factor in. Last but not least, following the notations of \cite{nelson2006use}, the original formulation of PIT method involves the sum of $Q$ products of $ n_i $ probability densities, which are inside the logarithm function so that those multiplications can not further be converted into summations of logarithm. See the last equation in Section~$3$ of \cite{nelson2006use}. Therefore, compared to the PIT method, the proposed methods are less sensitive to numerical issues, and is numerically  more stable especially when $ n_i $ is large. 

To our best knowledge, we are not aware of an implementation of the PIT method in either Python or R. Hence, we implemented it according to the original formulation. The only enhancement was to take natural logarithm of the last equation in Section~$3$ of \cite{nelson2006use}, otherwise it will quickly lead to numerical issues when optimizing it. With a slight abuse of notations (only in this section), following the notations of \cite{nelson2006use}, we solve the following optimization problem  

\begin{equation} \label{eq:pit}
\min_{\boldsymbol{\beta}, \boldsymbol{\theta}} \sum_{i=1}^{N} \left( -\ln\left( \sum_{q=1}^{Q} \prod_{k=1}^{n_i} f(y_{ik}|x_{ik}, F^{-1}_{\theta}(\Phi(d_q)), \boldsymbol{\beta}) \phi(d_q) w_q\right) \right).  
\end{equation}
The number of points used to approximate integrals is either $ 2 $ or $ 4 $, i.e., $ Q = 2, 4 $ to keep a balance between accuracy and computing time in this section, and the values of $ z_q, \eta_q, q=1, \ldots, Q$ are found in Table $25.10$ of \cite{abramowitz1988handbook}. Likewise, we assume SDTN distributions on the random effects with a random intercept only model. The modeling results for $ n=300, 500 $ with $ p=3 $ are reported in Table \oldref{tab:intercept_pit}. 
 
 \begin{table}[htbp!]
 	\begin{center}
 		\caption {Comparisons between the PIT method and the proposed methods}  \label{tab:intercept_pit}
 	\resizebox{\textwidth}{!}{\begin{tabular}{|c|c|cccc|cccc|} \hline
 		&	&  \multicolumn{4}{|c|}{$n=300$} & \multicolumn{4}{|c|}{$n=500$}  \\ \hline 
 		& True   & PIT ($Q=2$)   & PIT ($Q=4$)   & PLS   & PRLS  & PIT ($Q=2$)   & PIT ($Q=4$)   & PLS   & PRLS  \\ \hline 
 		$ \beta_{1, 0} = \beta_0 + \gamma_{1, 0} $ & $0.000$ & $0.000$ & $0.000$ & $0.000$ & $0.000$ & $0.000$ & $0.000$ & $0.000$ & $0.000$ \\ \hline 
 		$ \beta_{2, 0} = \beta_0 + \gamma_{2, 0}$  & $0.144$ & $0.000$ & $0.000$ & $0.066$ & $0.065$ & $0.000$ & $0.000$ & $0.013$ & $0.020$ \\ \hline 
 		$ \beta_1 $  & $1.000$      & $1.052$ & $1.052$ & $1.044$ & $1.084$ & $0.990$ & $1.512$ & $0.987$ & $0.985$ \\ \hline 
 		$ \beta_2 $  & $1.000$      & $0.995$ & $0.995$ & $0.990$ & $0.955$ & $1.014$ & $1.061$ & $1.018$ & $1.015$ \\ \hline 
 		$ \sigma $  & $1.000$      & $0.976$ & $0.976$ & $0.983$ & $0.993$ & $1.007$ & $1.144$ & $1.000$ & $1.014$ \\ \hline 
 		RMSE   &        & $0.069$ & $0.069$ & $0.041$ & $0.056$ & $0.065$ & $0.248$ & $0.060$ & $0.057$ \\ \hline  
 	\end{tabular}}
\end{center}
 \end{table} 
  
We did consider $n=1000$ as well but the numerical instability arises as the multiplication of many densities leads to exactly $ 0 $ before taking the natural logarithm in equation (\oldref{eq:pit}). Therefore, the results for $ n=1000 $ have to be dropped. Actually, even with $ n=500 $, the results for $ Q=4 $ are much worse than others as shown in Table~\oldref{tab:intercept_pit} due to the same issue. Also note that since the PIT method estimates $ \beta_0 $ less than $ 10^{-5} $, $s(\gamma_{i, 0}) = \sqrt{\eta^2(1-\frac{2\rho\phi(\rho)}{2\Phi(\rho)-1})} $ is not available since $ \rho =\beta/\eta$ is too close to $ 0 $ rendering $ \frac{2\rho\phi(\rho)}{2\Phi(\rho)-1} $ as a $ 0/0 $ indeterminate type that leads to NaN (Not a Number) in the Python programming language. Hence, the RMSE reported in Table \oldref{tab:intercept_pit} is only based on $ 5 $ parameters $ \beta_{1, 0}, \beta_{2, 0}, \beta_1, \beta_2, \sigma $ to ensure a fair comparison. It is apparent from Table~\oldref{tab:intercept_pit} that the overall RMSE of the proposed methods are lower than that of the PIT method, for example, with $ n=300 $, the RMSE for PLS and PRLS is $ 0.041, 0.056 $, respectively, contrasting to $ 0.069 $ for the PIT method. In summary, the proposed methods are not only numerically more stable, but also yield estimates that are closer to the true parameters.

\section{Real-World Applications} \label{sec:app}
\subsection{The Discounted Sales Data}
Let us revisit the motivating example introduced in Section \oldref{sec:intro}. The traditional LME model is included for comparison purposes, and it was fitted using the \textit{lme4} package in R. The model results of PLS, PRLS and \textit{lme4} are reported in Table~\oldref{tab:app3} for the estimated fixed coefficients. 

\begin{table}
	\centering 
	\caption{Model results for the discounted sales example.} \label{tab:app3}
	\begin{tabular}{|c|c|c|c|}  \hline 
		& \multicolumn{1}{c|}{\textit{lme4}} & \multicolumn{1}{c|}{PLS} & \multicolumn{1}{c|}{RPLS}   \\ \hline
		$ \beta_0 $ & $0.106$                    & $0.164$                   & $0.164$               \\ \hline 
	 $ \beta_1 $ & $-0.319$                   & $0.259$                   & $0.270$                \\ \hline
	 S.D. of random intercept   & $0.671$                    & $0.095$                   & $0.094$                \\ \hline
	 S.D. of random slope   & $0.361$                    & $0.075$                   & $0.156$                 \\ \hline
	 S.D. of residuals & $1.388$                    & $1.445$                   & $1.447$                 \\ \hline \hline
	Marginal $R^2$   & $0.000$                    & $0.000$                   & $0.000$                 \\ \hline
	Conditional $ R^2 $   & $0.190$                    & $0.488$                   & $0.606$                 \\ \hline
	\end{tabular}
\end{table}

From Table \oldref{tab:app3}, it is very apparent that both the proposed models produce non-negative $ \hat{\beta}_1 $ as compared to $-0.319$ of the traditional LME. With the proposed methods, heuristics for correcting the ``wrong'' signs are no longer needed, and the modeling results can be applied directly in practice. We also note that the conditional $ R^2 $ of the proposed methods are higher than that of the \textit{lme4} method, although all of the three methods have a $ 0 $ marginal $ R^2 $ meaning the random effects explain all of the variation in the data. The proposed methods not only preserve model interpretability and sign correctness, but also have better model fitting as measured by the conditional $R^2$. 

\begin{comment}
\begin{table}[htbp!]
\caption{Estimated overall regression coefficients of the Discount Rate.} \label{tab:app4}
\centering
\begin{tabular}{|c|c|c|c|} \hline \hline
Store Cluster	& lme4 & Saddlepoint with Uniform & Saddlepoint with Triangular \\ \hline
$A$ & -0.332 & 0.000  & 0.000   \\ \hline
$B$ & -0.332	& 0.000  & 0.000  \\ \hline
$C$ & -0.332	& 0.115  & 0.328  \\ \hline
$D$ & -0.332	& 0.000  & 0.000 \\ \hline
$E$ & -0.332	& 0.114  & 0.324 \\ \hline
$F$ & -0.332 & 0.000 & 0.000 \\ \hline \hline
\end{tabular}
\end{table}
\end{comment}
%From Table \oldref{tab:app4}, all of the estimated overall slopes are positive and are similar to each other for the proposed methods, while none of the results from \textit{lme4} is positive. 

\subsection{The Sleep Deprivation Study}
Consider a dataset on the reaction time per day for subjects in a sleep deprivation study \citep{bolker2013strategies}. This dataset is accessible in the \textit{lme4} package in R \citep{bates2014fitting}. There are $ 18 $ subjects in the dataset. On day $ 0 $, subjects had the normal amount of sleep, followed by the next $ 10 $ days when they were restricted to $ 3 $ hours of sleep per night only. The response variable is the reaction times in milliseconds.

\begin{figure}[htbp!]
	\hspace*{-1cm} \includegraphics[width=1.1\textwidth, height=3.8in]{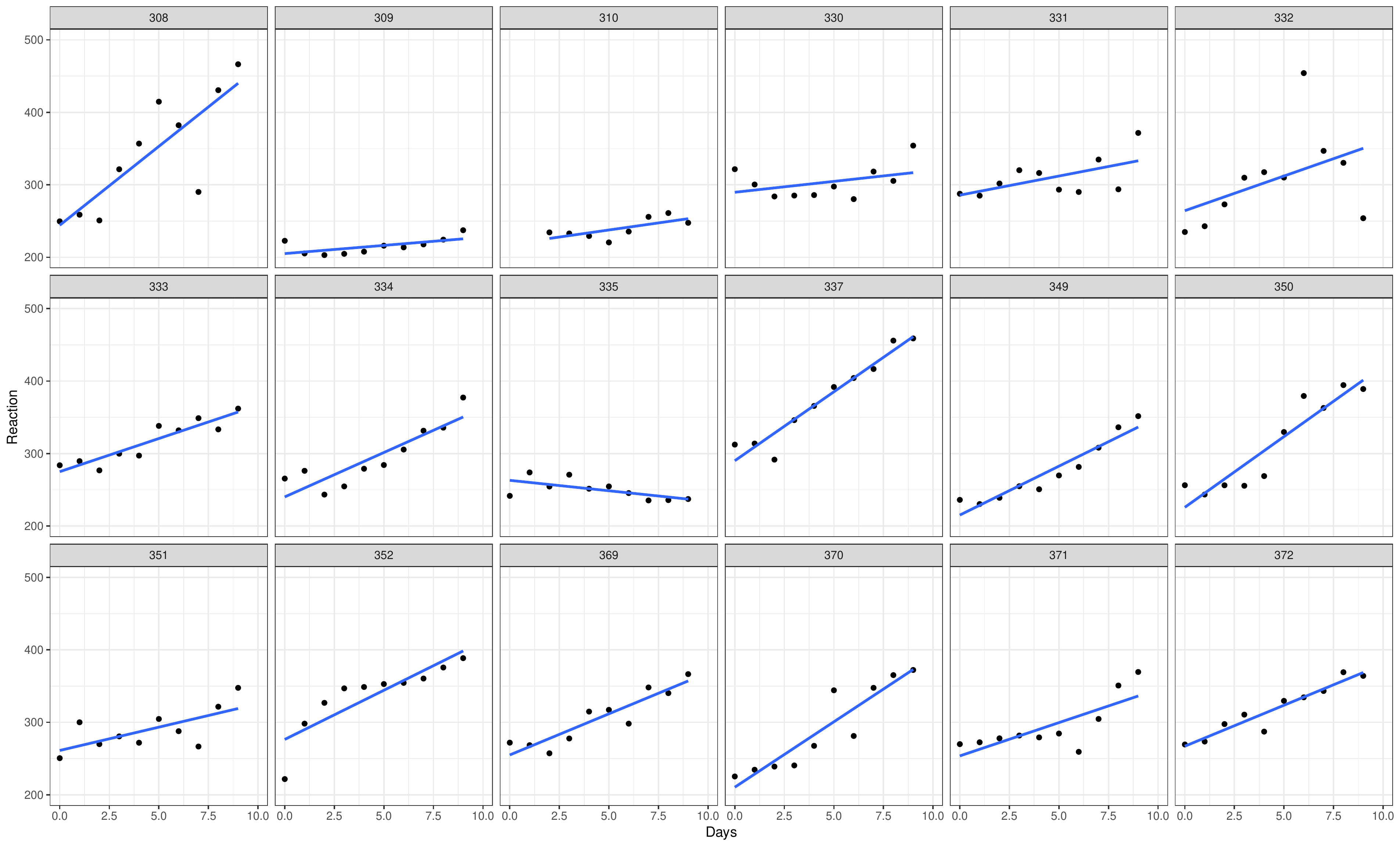}
	\caption{Scatter plots by subject.}
	\label{fig:app}
\end{figure}

Before we conduct any formal analysis, we plotted the data grouped by subjects as shown in Figure~\oldref{fig:app}. While most of the graphs match our intuition that the response time increases as the number of sleep-deprived days, we observe that the response curves across subjects are rather different. Thus, we use the full model that includes both the random intercept and random slope. We consider \textit{lme4}, PLS and PRLS and report the results in Table~\oldref{app:res}. From Table~\oldref{app:res}, the estimated regression coefficients are close for all the three methods considered: for example, the estimated coefficient for Days are $10.467, 10.789, 10.795$, respectively, for \textit{lme4}, PLS and PRLS. In terms of model fits, the proposed methods clearly outperform \textit{lme4} since the conditional $R^2$ for PLS and PRLS are $0.803$ and $ 0.817$ that are apparently higher than $0.702$ of \textit{lme4}. The same pattern holds for marginal $ R^2 $ as well.    

We also report the estimated random effects. We show the overall effect (fixed effect + random effect) of all $ 18 $ subjects in 
Table~\oldref{tab:app}. It is observed that the estimates are similar across the three methods for all subjects. In addition, the estimates are more consistent between PLS and PRLS than between PLS/PRLS and \textit{lme4}. This is not unexpected as the two proposed methods share more similarity than the \textit{lme4} method. Among all of the $18$ subjects, the most interesting individual is subject $ 335 $, where the estimated slope is negative from \textit{lme4}, while the estimates from the proposed methods are $ 0.000 $. This is because of the use of SDTN in the model specification which reflected the intuition that as the number of sleep-deprived days increases, the response time cannot decrease. The results from the proposed PLS and PRLS matches this intuition and should therefore be preferred over those from the unconstrained \textit{lme4}.
\begin{table}[htbp!]
	%\centering
	%\captionsetup{justification=centering}
	\caption{Model results for the sleep deprivation study.} \label{app:res}
	\centering
	\begin{tabular}{|l|c|c|c|}
	\hline
	& \textit{lme4}                        & PLS                         & PRLS                        \\ \hline
	\multicolumn{1}{|c|}{Intercept} & \multicolumn{1}{c|}{251.405} & \multicolumn{1}{c|}{250.389} & \multicolumn{1}{c|}{250.356} \\ \hline
	\multicolumn{1}{|c|}{Days}      & \multicolumn{1}{c|}{10.467}  & \multicolumn{1}{c|}{10.789}  & \multicolumn{1}{c|}{10.795}  \\ \hline
	S.D. of random intercept        & 25.051                       & 25.555                       & 26.407                       \\ \hline
	S.D. of random slope            & 5.988                        & 6.228                        & 6.232                        \\ \hline
	S.D. of residuals               & 25.565                       & 20.140                       & 19.509                       \\ \hline \hline
	Marginal $R^2$                      & 0.415                     & 0.467                     & 0.463                      \\ \hline
	Conditional $R^2$                      & 0.702                     & 0.803                     & 0.817                      \\ \hline
\end{tabular}
	%\end{adjustbox}
	%\end{center}
\end{table}

\begin{table}[htbp!]
	\caption{The overall effects (fixed effect + random effect) of the $ 18 $ subjects.} \label{tab:app}
	\begin{center}
		\begin{adjustbox}{max width=\textwidth}
		\begin{tabular}{|l|l|c|c|c|c|c|c|} \hline
			&     & Subject 308                 & Subject 309                & Subject 310                & Subject 330 & Subject 331 &
			Subject 332 \\ \hline \hline
			
			\multicolumn{1}{|c|}{\multirow{2}{*}{\textit{lme4}}} & \multicolumn{1}{c|}{Overall Intercept} & \multicolumn{1}{c|}{252.918} & \multicolumn{1}{c|}{211.031} & \multicolumn{1}{c|}{212.224} & 275.924      & 274.320 & 260.627 \\ \cline{2-8}
			
			\multicolumn{1}{|c|}{}                      & \multicolumn{1}{c|}{Overall Slope}                                                & \multicolumn{1}{c|}{19.791}   & \multicolumn{1}{c|}{1.868}   & \multicolumn{1}{c|}{5.079}  & 5.499       & 7.273 & 10.159       \\ \hline \hline
			
			\multirow{2}{*}{PLS}                        & Overall Intercept                                                                 & 244.621                      & 208.148                      & 202.232                      & 297.813      & 288.122 & 246.798      \\ \cline{2-8}
			
			& Overall Slope                                                                     & 21.580                        & 1.330                        & 6.512                        & 5.652       & 4.700 & 14.301       \\ \hline \hline
			
			\multirow{2}{*}{PRLS}                       & Overall Intercept                                                                 & 244.592                      & 208.110                      & 202.237                      & 297.743      & 288.107 & 246.910      \\ \cline{2-8}
			
			& Overall Slope                                                                     & 21.581                        & 1.340                        & 6.503                        & 5.672       & 4.688 & 14.256       \\ \hline \hline \hline \hline

			&     & Subject 333                 & Subject 334                & Subject 335                & Subject 337 & Subject 349 &
			Subject 350 \\ \hline \hline
			
			\multicolumn{1}{|c|}{\multirow{2}{*}{\textit{lme4}}} & \multicolumn{1}{c|}{Overall Intercept} & \multicolumn{1}{c|}{268.561} & \multicolumn{1}{c|}{243.953} & \multicolumn{1}{c|}{251.984} & 286.173      & 225.651 & 237.540 \\ \cline{2-8}
			
			\multicolumn{1}{|c|}{}                      & \multicolumn{1}{c|}{Overall Slope}                                                & \multicolumn{1}{c|}{10.180}   & \multicolumn{1}{c|}{11.583}   & \multicolumn{1}{c|}{-0.439}  & 19.095       & 11.748 & 17.224       \\ \hline \hline
			
			\multirow{2}{*}{PLS}                        & Overall Intercept                                                                 & 275.692                      & 247.760                      & 253.771                      & 292.131      & 220.663 & 229.012      \\ \cline{2-8}
			
			& Overall Slope                                                                     & 8.881                        & 10.031                        & 0.000                        & 18.323       & 11.898 & 18.401       \\ \hline \hline
			
			\multirow{2}{*}{PRLS}                        & Overall Intercept                                                                 & 274.682                      & 247.612                      & 253.777                      & 292.071      & 220.612 & 228.478      \\ \cline{2-8}
			
			& Overall Slope                                                                     & 8.912                        & 10.051                        & 0.000                        & 18.350       & 11.910 &  18.451       \\ \hline \hline \hline \hline

			&     & Subject 351                 & Subject 352                & Subject 369                & Subject 370 & Subject 371 &
			Subject 372 \\ \hline \hline
			
			\multicolumn{1}{|c|}{\multirow{2}{*}{\textit{lme4}}} & \multicolumn{1}{c|}{Overall Intercept} & \multicolumn{1}{c|}{256.321} & \multicolumn{1}{c|}{272.334} & \multicolumn{1}{c|}{254.664} & 224.929      & 252.311 & 263.827 \\ \cline{2-8}
			
			\multicolumn{1}{|c|}{}                      & \multicolumn{1}{c|}{Overall Slope}                                                & \multicolumn{1}{c|}{7.392}   & \multicolumn{1}{c|}{13.979}   & \multicolumn{1}{c|}{11.340}  & 15.451       & 9.462 & 11.726       \\ \hline \hline
			
			\multirow{2}{*}{PLS}                        & Overall Intercept                                                                 & 265.060                      & 268.932                      & 257.112                      & 212.870      & 261.842 & 267.758      \\ \cline{2-8}
			
			& Overall Slope                                                                     & 5.447                        & 15.921                        & 10.666                        & 17.289       & 6.947 & 11.101       \\ \hline \hline
			
			\multirow{2}{*}{PRLS}                        & Overall Intercept                                                                 & 265.030                      & 269.541                      & 257.084                      & 212.810      & 261.749 & 267.778      \\ \cline{2-8}
			
			& Overall Slope                                                                     & 5.513                        & 15.848                        & 10.689                        & 17.292       & 6.970 & 11.117       \\ \hline \hline \hline \hline			
		\end{tabular}
	\end{adjustbox}
	\end{center}
\end{table}

\section{Concluding Remarks} \label{sec:conclusions} 
In this paper, we work under the framework of linear mixed effects model. We assume the SDTN distribution on the random effects instead of the Normal distribution to impose sign constraints on the overall effects in a theoretically sound way. This change has profound impact on the estimation methods because the exact distribution of $ \boldsymbol{y} $  becomes analytically intractable. We lay a solid foundation by establishing properties of a SDTN distribution and then proposed two methods: PLS and PRLS for estimating the unknown model parameters. Both the simulation studies and the application examples show their satisfactory performance as compared to the unconstrained \textit{lme4}. When there is practical justification or domain knowledge to impose sign constraints on the overall regression parameters, the proposed methods work best in finding alternative solutions that allow intuitive interpretation of the results. 

We discuss a few future extensions motivated by this research. A natural extension of this research is to consider the generalized linear mixed effects model (GLMM) so that it can be applied to broader types of data such as binary or count outcomes. The framework of the linear mixed effects model is sufficient for our current practical needs, but GLMM has broader applications in social and economic research, medical studies, and the pharmaceutical industry.  

%\section*{Disclosure statement}
%No potential conflict of interest was reported by the authors.
%\vspace*{-0.3cm}
%\section*{Funding}
%The research is wholly supported by Precima, a Nielsen company, an industry-leading provider of powerful retail data research, analytics and applications, and is part of the Nielsen Connect family. All authors are senior research fellows employed by Nielsen.   

\nocite{zolotarev1967generalization}

%\bibliographystyle{tfnlm}
%\bibliography{Paper}
\bibliographystyle{apacite}
\bibliography{sample}
%\newpage
%\newpage
%\vspace*{-0.3cm}
%\begin{comment}
%\vspace*{-0.3cm}
\appendix

\newpage
\pagestyle{empty}
\clearpage
\pagestyle{plain} 
\pagenumbering{arabic}
\begin{center}
	\Large \textbf{Supplemental Document}
\end{center}

\section*{Section A. Proof of Lemma 3.2} \label{append:A}
Lemma 3.2 (i) and (ii) are obvious from properties of standard Normal distribution PDF $\phi(\xi)$. To prove (iii), notice that $2\Phi(\rho) > 1$ for all $\rho>0$. Since $\phi(\rho)>0$, we have
\begin{equation} \nonumber
\frac{2\rho \phi(\rho)}{2\Phi(\rho)-1} > 0, \,\, \forall \rho>0.
\end{equation}
Therefore,
\begin{equation} \nonumber
\var[x] \, = \, \eta^2 \left[1 - \frac{2\rho \phi(\rho)}{2\Phi(\rho)-1}\right] \, \leq \, \eta^2.
\end{equation}
To prove (iv), we examine the following function
\begin{equation}
g(\rho) \, \triangleq \, \frac{2\rho \phi(\rho)}{2\Phi(\rho)-1}.
\end{equation}
It is clear this function is differentiable on $(0,+\infty)$. Noticing that the derivate of $\phi(\rho)$, $\phi'(\rho) = - \rho \phi(\rho)$, the derivative of $g(\rho)$ is written as
\begin{eqnarray}
g'(\rho) & = & \frac{2\phi(\rho) + 2\rho \phi'(\rho)}{2\Phi(\rho)-1} - \frac{4\rho [\phi(\rho)]^2}{[2\Phi(\rho)-1]^2} \nonumber\\
& = & \frac{[2\phi(\rho) + 2\rho \phi'(\rho)][2\Phi(\rho)-1]-4\rho [\phi(\rho)]^2}{[2\Phi(\rho)-1]^2} \nonumber\\
& = & \frac{2\phi(\rho)\left[ (1-\rho^2)(2 \Phi(\rho)-1) - 2\rho \phi(\rho)\right]}{[2\Phi(\rho)-1]^2} \nonumber
\end{eqnarray}
We further let
\begin{equation} \nonumber
t(\rho) \, \triangleq \, (1-\rho^2)(2 \Phi(\rho)-1) - 2\rho \phi(\rho).
\end{equation}
It is clear that $t(\rho)$ is continuous in $\rho$ and $t(0)=0$. We also have
\begin{equation} \nonumber
t'(\rho) \, = \, -2\rho(2\Phi(\rho)-1) < 0,
\end{equation}
for all $\rho>0$. Therefore, $t(\rho)<0$ for all $\rho>0$. It implies that $g'(\rho)<0$ for all $\rho>0$. Therefore, $g(\rho)$ is a monotonically decreasing function on $(0,\infty)$. Hence (iv) holds readily. For (v), it is straightforward to verify that the PDF of $x'$ is given by
\begin{eqnarray}
f(x^{\prime}) &=&
\left\{
\begin{array}{ll}
\frac{1}{k_1} f\left(\frac{x - k_0}{k_1}\right) & \mbox{if } k_1 > 0 \\[5pt]
-\frac{1}{k_1} f\left(\frac{x - k_0}{k_1}\right) & \mbox{if } k_1 < 0 \\[5pt]
\end{array}
\right. \nonumber\\
&=& \frac{1}{|k_1|} \frac{1}{\eta} \frac{\phi(  \frac{x-k_0 - 0}{k_1\eta})}{2\Phi(\rho) - 1} \nonumber \\
&=& \frac{1}{|k_1| \eta} \frac{\phi(\frac{x-k_0}{k_1\eta})}{2\Phi(\rho) - 1} \nonumber
\end{eqnarray}
The last equation is the PDF of $ \mathcal{DTN}(k_0, k_1^2\eta^2, \rho)  $. \qed

\section*{Section B. Proof of Theorem 3.3} \label{append:B}
In this proof, we will use the well known Lindeberg-Feller theorem \citep{zolotarev1967generalization}:
\textit{
	%\begin{theorem}
	Suppose that $x_1,x_2,\cdots$ are independent random variables such that $\Ebld[x_i] =\mu_i$ and $\var[x_i] = \sigma_i^2 < +\infty$ for all $i=1,2,\cdots$. Define:
	\begin{eqnarray}
	y_i & = & x_i - \mu_i, \nonumber \\
	s_n^2 & = & \sum_{i=1}^n \var\left[y_i\right] \, = \, \sum_{i=1}^n \sigma_i^2. \nonumber
	\end{eqnarray}
	If the Lindeberg condition
	\begin{equation} \label{eq:linderberg}
	\mbox{for every }\epsilon>0, \frac{1}{s_n^2} \sum_{i=1}^n \Ebld\left[y_i^2 \cdot \mathbf 1_{|y_i|\geq \epsilon s_n}\right] \, \rightarrow \, 0 \,\,\mbox{as } n\rightarrow \infty
	\end{equation}
	is satisfied, then 
	\begin{equation} \nonumber
	\frac{\sum_{i=1}^{n}\left(x_i - \mu_i\right)}{s_n} \, \xrightarrow{d} \, \mathcal N(0,1).
	\end{equation}
}
For Theorem~$3.3$ to hold, it suffices to verify the Lindeberg condition (\oldref{eq:linderberg}). First, by Lemma~$3.2$, item (iii), we have
\begin{equation} \nonumber
\var[y_i]\leq \eta_i^2 < \infty.
\end{equation}
Next, since $\rho_i \geq \underline{\rho}$ for each $i=1,2,\cdots$, by Lemma~$3.2$ item (iv), we have
\begin{equation} \nonumber
\var[y_i] \, \geq \, \eta_i^2 \left[1 - \frac{2\underline{\rho}\phi(\underline{\rho})}{2\Phi(\underline{\rho})-1}\right] \, \geq \, \underline\eta^2 \left[1 - \frac{2\underline{\rho} \phi(\underline{\rho})}{2\Phi(\underline{\rho})-1}\right] \, \triangleq \, v^2,
\end{equation}
where $  v = \underline{\eta} \sqrt{1 - \frac{2\underline{\rho} \phi(\underline{\rho})}{2\Phi(\underline{\rho})-1}} > 0 $. It follows that $s_n^2 \geq n v^2$ for all $n=1,2,\cdots$. By Lemma~$3.2$,
$y_i \, \sim \, \mathcal{DTN}(0,\eta_i,\rho_i)$. Therefore, with $ u_i \triangleq \frac{y_i}{\eta_i}$, for any given $\epsilon>0$ and for each $i=1,2,\cdots$, we have
\begin{eqnarray}
\Ebld\left[ y_i^2 \cdot \mathbf 1_{ |y_i|>\epsilon s_n }\right] & = & \int_{-\rho_i \eta_i}^{\rho_i \eta_i} y_i^2 f_{\mathcal{DTN}}\left(y_i;0, \eta_i, \rho_i\right) \cdot \mathbf 1_{ |y_i|>\epsilon s_n } d y_i \nonumber \\
& = & \left\{
\begin{array}{ll}
0 & \mbox{if } \epsilon s_n \geq \rho_i\eta_i \\[5pt]
2 \displaystyle {\int_{\epsilon s_n}^{\rho_i\eta_i} y_i^2 f_{\mathcal{DTN}}\left(y_i;0, \eta_i, \rho_i\right) d y_i }& \mbox{if } \epsilon s_n < \rho_i\eta_i
\end{array}
\right. \nonumber\\
& \leq & 2 \int_{\epsilon s_n}^{\infty}  y_i^2 \frac{1}{\eta_i (2 \Phi(\rho_i)-1)} \phi\left(\frac{y_i}{\eta_i}\right) d{y_i}\nonumber\\
&=& \frac{2}{\eta_i} \frac{1}{2\Phi(\rho_i) - 1} \frac{1}{\sqrt{2\pi}}
\int_{\epsilon s_n}^{\infty} {y_i}^2 \exp{\left(-\frac{y_i^2}{2\eta_i^2}\right)} d y_i \nonumber \\
&=& \frac{2}{\eta_i} \frac{1}{2\Phi(\rho_i) - 1} \frac{1}{\sqrt{2\pi}}
\left(\eta_i^3
\int_{\frac{\epsilon s_n}{\eta_i}}^{\infty} u_i^2 \exp\left(-\frac{u_i^2}{2}\right) du_i\right) \nonumber \\
&=& \eta_i^2 \frac{1}{2\Phi(\rho_i) - 1} \sqrt{\frac{2}{\pi}}
\int_{\frac{\epsilon s_n}{\eta_i}}^{\infty} u_i^2 \exp\left(-\frac{u_i^2}{2}\right) du_i \nonumber \\
&\leq& \overline \eta^2 \frac{1}{2\Phi(\underline \rho) - 1} \sqrt{\frac{2}{\pi}}
\int_{\frac{\epsilon \sqrt n v}{\overline \eta}}^{\infty} u_i^2 \exp{\left(-\frac{u_i^2}{2}\right)} du_i. \nonumber
\end{eqnarray}
Notice that
\begin{equation} \nonumber
\int u_i^2 \exp{\left(-\frac{u_i^2}{2}\right)} du_i = \sqrt{\frac{\pi}{2}} \text{erf}\left(\frac{u_i}{\sqrt{2}}\right) - u_i\exp{\left(-\frac{u_i^2}{2}\right)}.  
\end{equation}
Hence, we have
\begin{equation} \nonumber
%\hspace*{-0.7cm}
\left. \begin{aligned}
\lim_{n\rightarrow \infty} &\frac{1}{s_n^2} \sum_{i=1}^n \Ebld[y_i^2 \cdot \mathbf 1_{|y_i|\geq \epsilon s_n}] \nonumber\\
& \leq  \lim_{n\rightarrow \infty}\frac{1}{s_n^2} \sum_{i=1}^n \left[\overline \eta^2 \frac{1}{2\Phi(\underline \rho) - 1} \sqrt{\frac{2}{\pi}}
\int_{\frac{\epsilon \sqrt n v}{\overline \eta}}^{\infty} u_i^2 \exp\left(-\frac{u_i^2}{2}\right) d u_i \right] \nonumber\\
& \leq  \lim_{n\rightarrow \infty} \frac{\overline \eta^2 }{n v^2} \frac{1}{2\Phi(\underline \rho) - 1} \sqrt{\frac{2}{\pi}} \sum_{i=1}^n \left[ \sqrt{\frac{\pi}{2}} \left(1-\mbox{erf}\left(\frac{\epsilon \sqrt n v}{\overline \eta}\right) \right) + \frac{\epsilon \sqrt n v}{\overline \eta} \exp\left(-\frac{\epsilon^2 n v^2}{2 \overline \eta^2}\right) \right]\nonumber \\
& =  \lim_{n\rightarrow \infty} \frac{\overline \eta^2 }{v^2} \frac{1}{2\Phi(\underline \rho) - 1} \sqrt{\frac{2}{\pi}} \left[ \sqrt{\frac{\pi}{2}} \left(1-\mbox{erf}\left(\frac{\epsilon \sqrt n v}{\overline \eta}\right) \right) + \frac{\epsilon \sqrt n v}{\overline \eta} \exp\left(-\frac{\epsilon^2 n v^2}{2 \overline \eta^2}\right) \right] \nonumber \\
& =  0, \nonumber
\end{aligned} \right.
\end{equation}
where the last equality is due to the fact that since $ \epsilon > 0 $ is finite, and $ v, \overline{\eta} $ are fixed constants,
\begin{equation} \nonumber
\lim\limits_{n \rightarrow \infty} \frac{\epsilon \sqrt{n} v}{\overline{\eta}} \rightarrow \infty \, \, \mbox{and} \,\,
\lim_{n\rightarrow \infty} \mbox{erf}\left(\frac{\epsilon \sqrt n v}{\overline \eta}\right) \, = \, 1 \, \, \mbox{and} \,\, \lim_{z\rightarrow \infty } z \exp\left(-\frac{z^2}{2}\right) \, = \, 0.  \qed
\end{equation}

\section*{Section C. Additional Figures in Simulation} %

\begin{figure}[htbp!]
	%\centering
	\hspace*{-0.7cm}
	\subfloat{{\includegraphics[width=0.5\linewidth, height=5.85cm]{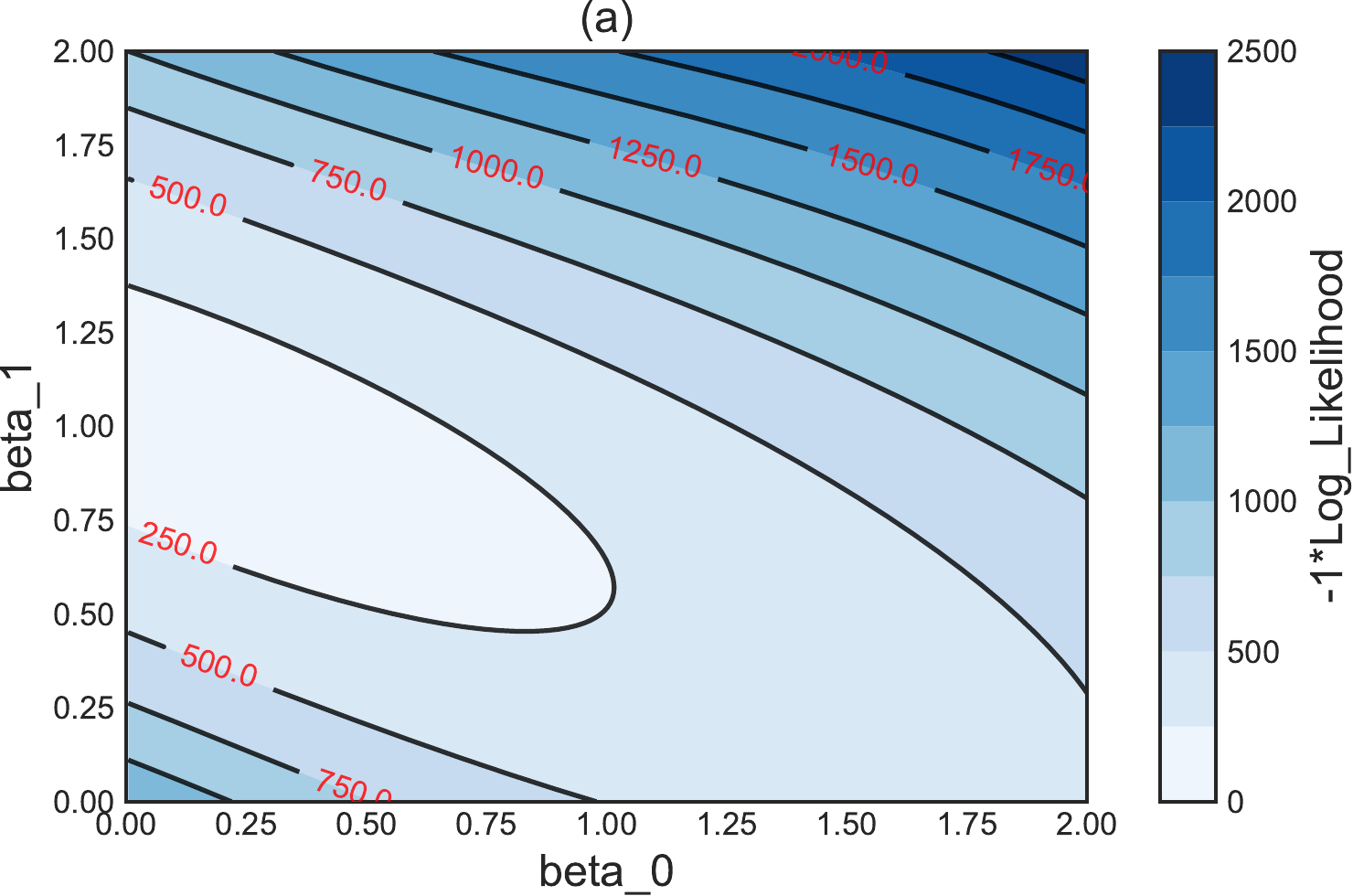}}}%
	\qquad
	\subfloat{{\includegraphics[width=0.5\linewidth, height=5.85cm]{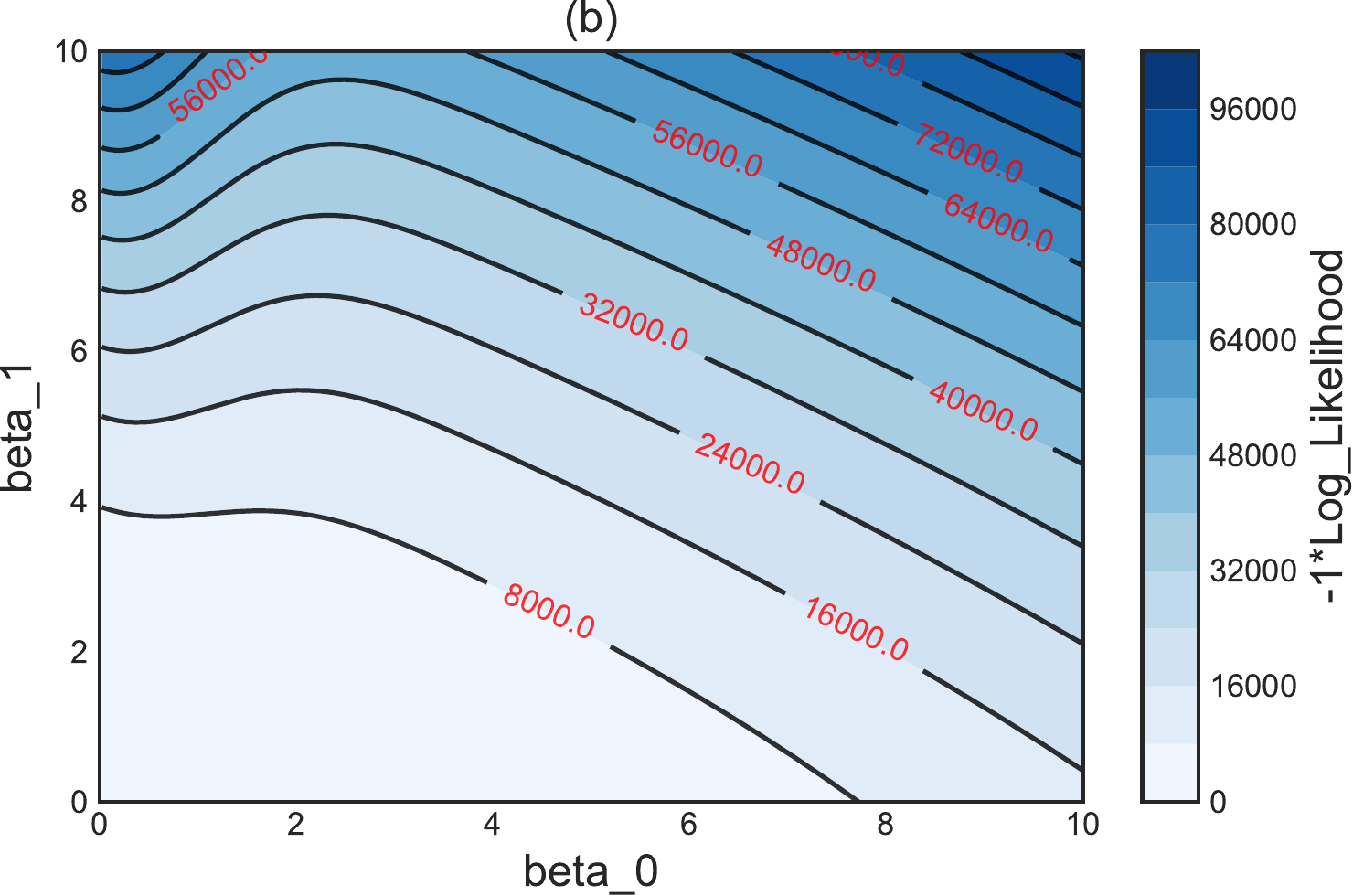}}}%
	\caption{Contour plots of the log-likelihood function in equation (15) viewed as a function of $\beta_0$ and $ \beta_1 $. The data range for (a) is $ 0 $ to $ 2 $, while the range for (b) is $0$ to $10$.}%
	\label{fig:con3}
\end{figure}

\begin{figure}[htbp!]
	%\centering
	\hspace*{-0.7cm}
	\subfloat{{\includegraphics[width=0.5\linewidth, height=5.85cm]{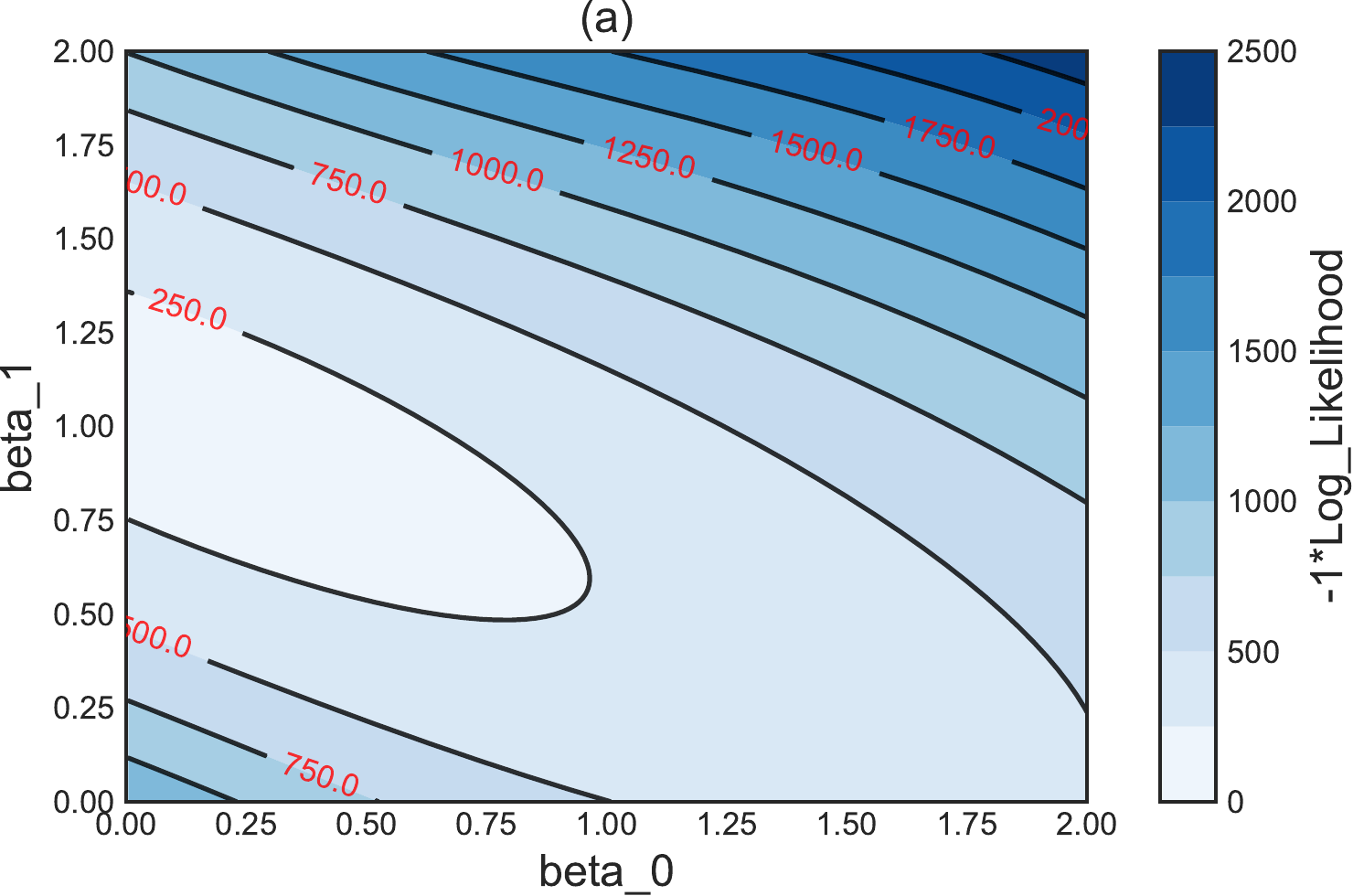}}}%
	\qquad
	\subfloat{{\includegraphics[width=0.5\linewidth, height=5.85cm]{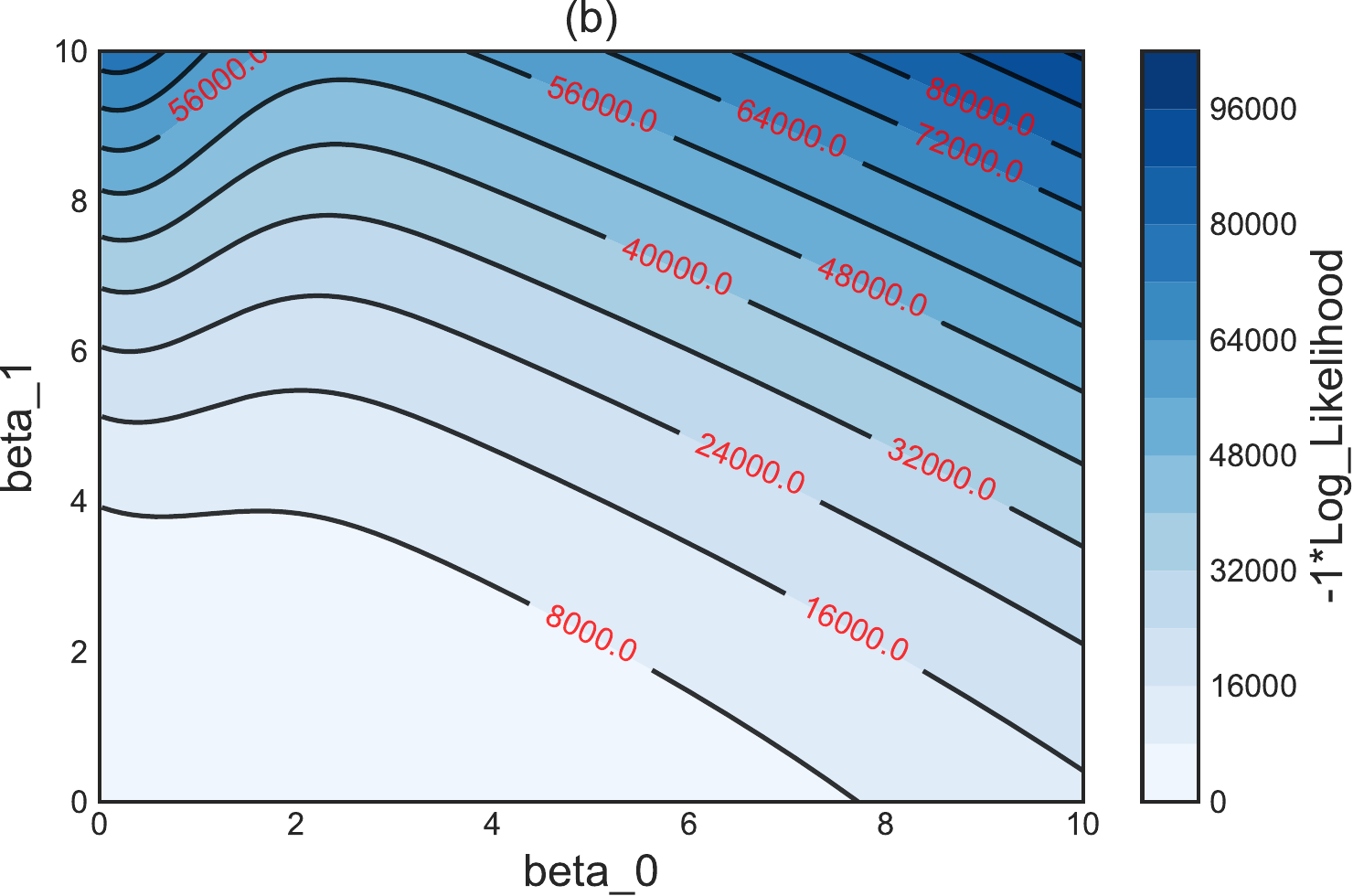}}}%
	\caption{Contour plots of the log-likelihood function in equation (16) viewed as a function of $\beta_0$ and $ \beta_1 $. The data range for (a) is $ 0 $ to $ 2 $, while the range for (b) is $0$ to $10$.}%
	\label{fig:con4}%
\end{figure}

\end{document}